\documentclass[preprint,12pt]{elsarticle}
\usepackage{amsmath}
\usepackage{graphicx,subcaption}
\usepackage{stackengine}
\usepackage{amssymb}
\usepackage{stmaryrd}
\usepackage{mathabx}
\usepackage{ulem}
\usepackage{color}

\journal{Acta Astronautica}

\begin{document}

\begin{frontmatter}

\title{Dynamical properties of the Molniya satellite
constellation: long-term evolution of orbital eccentricity}


\author{Elisa Maria Alessi\corref{cor}$^{(1,2)}$}
\cortext[cor]{Corresponding author}
\ead{elisamaria.alessi@cnr.it}
\author{Alberto Buzzoni$^{(3)}$, J\'er\^ome Daquin$^{(4)}$, Albino Carbognani$^{(3)}$ \& Giacomo Tommei$^{(5)}$}
\address{
$^{(1)}$IMATI-CNR, Istituto di Matematica Applicata e Tecnologie Informatiche ``E. Magenes'', 
Via Alfonso Corti 12, 20133 Milano, Italy\\
$^{(2)}$IFAC-CNR, Istituto di Fisica Applicata ``N. Carrara'', Via Madonna del Piano 10, 50019 Sesto Fiorentino
(FI), Italy\\
$^{(3)}$INAF-OAS, Osservatorio di Astrofisica e Scienza dello Spazio, Via P. Gobetti 93/3 40129 Bologna, Italy\\
$^{(4)}$naXys, Department of Mathematics, University of Namur, 8 rempart de la Vierge, 5000 Namur, Belgium \\
$^{(5)}$Universit\`a di Pisa, Dipartimento di Matematica, Largo B. Pontecorvo 5, 56127 Pisa, Italy\\
}

\begin{abstract}
The aim of this work is to analyze the orbital evolution of the mean eccentricity given by the Two-Line Elements (TLE) set of the Molniya satellites constellation. The approach is bottom-up, aiming at a synergy between the observed dynamics and the mathematical modeling. Being the focus the long-term evolution of the eccentricity, the dynamical model adopted is a doubly-averaged formulation of the third-body perturbation due to Sun and Moon, coupled with the oblateness effect on the orientation of the satellite. The numerical evolution of the eccentricity, obtained by a two-degree-of-freedom model assuming different orders in the series expansion of the third-body effect, is compared against the mean evolution given by the TLE. The results show that, despite being highly elliptical orbits, the second order expansion catches extremely well the behavior. Also, the lunisolar effect turns out to be non-negligible for the behavior of the longitude of the ascending node and the argument of pericenter. The role of chaos in the timespan considered is also addressed. Finally, a frequency series analysis is proposed to show the main contributions that can be detected from the observational data.
\end{abstract}


\end{frontmatter}

\parindent=0.5 cm


\begin{figure*}[!t]
\centering
\includegraphics[width=0.8 \hsize]{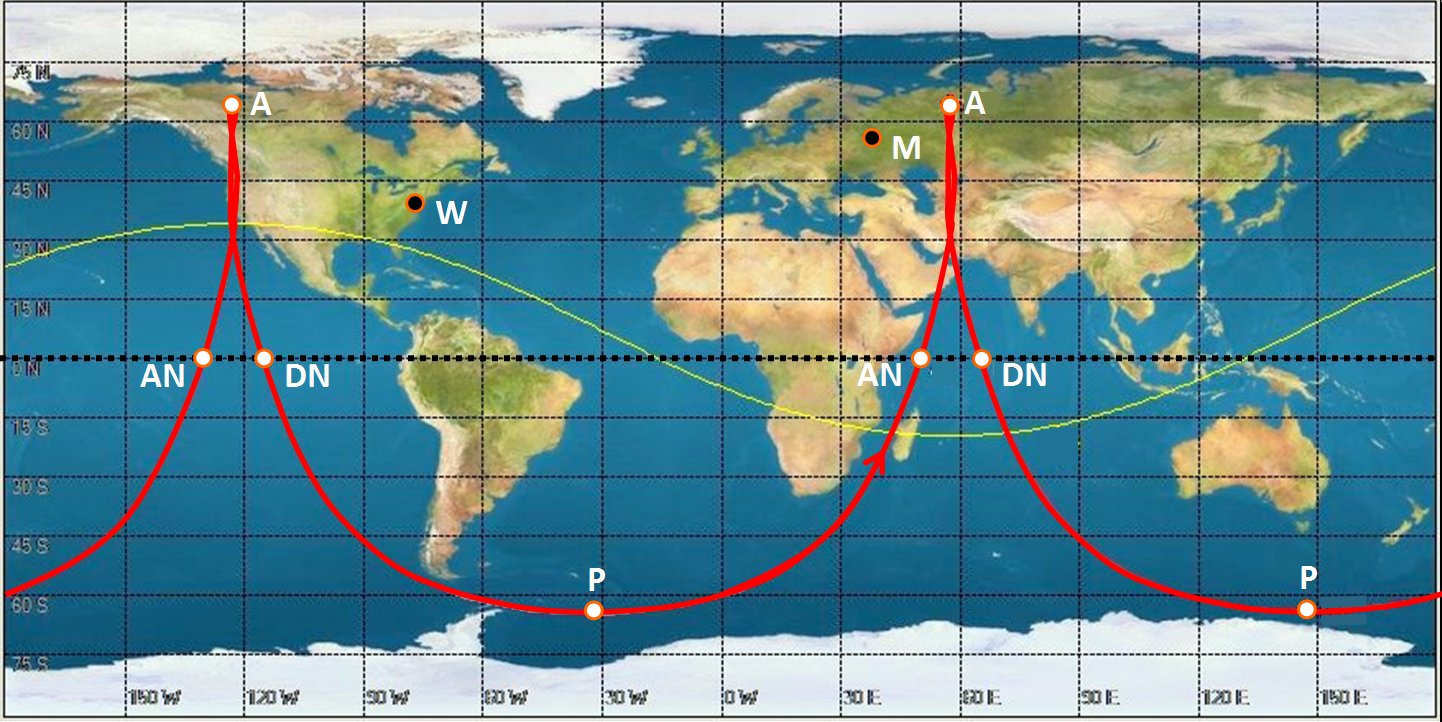}
\caption{An illustrative example of Molniya orbit ground track. Two 12-hr orbits are displayed to
span a full day. The nominal orbital parameters
are assumed, for illustrative scope, according to the template given in \cite{fortescue95}.
In particular, for our specific choice, we set the relevant geometric orientation parameters 
$(i,\omega) = (63.4, 270)$  deg, with orbit scale length parameters (in km) $(a,h_p,h_a)_{km} = 
(26560, 1000, 39360)$. This implies an eccentricity $e = 0.72$ and a period $P = 720$~min.
Note the extremely asymmetric location of the 
ascending (``AN'') and descending (``DN'') nodes, due to the high eccentricity of the orbit,
and the perigee (``P''), always placed in the Southern
hemisphere. Along the two daily apogees (``A''), the satellite hovers first Russia and then Canada/US. 
The visibility horizon (aka the ``footprint'') attained by the Molniya at its apogee over Russia, 
is above the displayed yellow line.}
\label{f01}
 \end{figure*}

\section{Introduction}
Starting in the mid-'60s, the Molniya program (the name standing for the Russian word ``Lightning'') 
inaugurated the innovative concept of ``satellite constellation''. For the first time,
indeed, a reliable communication service for military and civilian applications was set in place 
by Soviet Union across the extended country territories through the coordinated action of a 
spacecraft network instead of relying on individual space 
relays (see, e.g., \cite{ulybyshev09,trishchenko11,ulybyshev15} for recent formal settlements of the general astrodynamical problem).

This combined strategy is nowadays a fully established way to approach space servicing worldwide, especially
in the Low Earth Orbit regime, in order to enable a set of short-period spacecraft to provide ground end-users 
with uninterrupted and reliable up/downlink as for high-quality telephony and remote-sensing surveys.

The successful performance of the Molniya program also opened to further applications, 
especially from the Soviet/Russian side, leading in the early '70s to a military constellation of 
Cosmos spacecraft (the so-called OKO system) on a ``Molniya orbit'' (so christened after
the name of the satellites) for early-warning detection of hostile ballistic missiles \cite{podvig02}.
In addition, since the early 2000s, a follow-up to the original Molniya program 
was pursued by CSI with the Meridian constellation \cite{zak20}, currently consisting of 8 spacecraft 
in a similar Highly Eccentric Orbit (HEO), more explicitly dedicated to military communication.
A few cases of Molniya-like satellites might also be recognized among American satellites for military
applications, in support to the Space-Based Infrared System (SBIRS). 
Overall, a total of some 150 objects can be recognized in a Molniya orbital regime from ground 
surveillance surveys \cite{mcgraw17,silha17}, although this might be a crude underestimate of the real
population of orbiting objects once the increasingly important fraction of space debris could be 
properly included.

An extensive observing campaign of the full Molniya constellation was carried out, between 2014 and 2017, 
by our group at Mexican and Italian telescopes 
\cite{buzzoni19,buzzoni20}. The accounted dataset actually included all the 43 satellites still in
orbit in 2014; since then, seven spacecraft have decayed so that, as of this writing, the survived Molniya 
constellation consists of only 36 satellites, namely, 35 in HEO and 1 in geostationary orbit (GEO).\footnote{Of these, one satellite, namely Molniya 1-S was placed
in geostationary orbit and it will therefore not be considered here.}
For all of them we actually deal with no-cooperant spacecraft. 

With this paper, the first of a series, we want to focus our analysis on the relevant astrodynamical properties of 
the Molniya satellite constellation
in principle through a ``heuristic'' approach to the problem. The historical records from the 
TLE database \cite{JSPOC/NORAD} are used as a reference, restraining our analysis to
all the data available up to January 1, 2019. We therefore rely on the past history of the survived spacecraft constellation (spanning a four-decades interval
as for the oldest satellites) to set the ``ground truth'' for an accurate ex-post dynamical analysis
of each object, aimed at singling out the selective action of the different physical players 
(i.e., lunisolar perturbations, geopotential, solar radiation pressure, atmospheric drag, etc.) that may modulate secular evolution of the orbit. In this work, we will focus especially on the behavior in eccentricity; the behavior of the semi-major axis will be treated separately in a following work.


\section{The Molniya orbit}

From a dynamical point of view, the adopted Molniya orbit offered a fully appealing alternative to
the GEO option (see \cite{richharia99, christopher09} for a comparative discussion).
A series of spacecraft routed along a HEO
much ``closer'' ($a \sim 26500$~km or equivalently an orbital period $P = 12$~hr) than GEO and inclined
($i \sim 63$ deg) orbital path \cite[e.g.,][]{fortescue95}, was actually better suited to cover
the high-latitude and wide-longitude extension of the Soviet
Union.\footnote{The former USSR (and present CSI republic) covered over
9600~km from east to west and over 4000~km from north to south, reaching up to
$83$N  deg in latitude.}

The solution adopted was chosen because satisfying the mandatory requirement of being in view of the Soviet territory as long as possible. Such an ``extended permanence'' over Russia is eased by a
HEO with a convenient choice of the argument of perigee (a
value of $\omega \sim 270$ deg appeared to be a best option) such as to raise
the satellite to nearly GEO altitude at its apogee when hovering Russia. To
safely maintain this configuration, however, the orbital plane has to be
twisted such as to counteract the effect of Earth gravitational dipole (the
so-called $J_2$ geopotential term) on the $\omega$ drift \cite[e.g.,][]{larson92}.
To set $\dot{\omega} \sim 0$, and thus ``freeze'' the line of apsides, one has
to incline the orbit by $i = \sin^{- 1} \left( 2 / \sqrt{5} \right)
\sim 63.4$ deg \cite{wiesel89}.

With these conditions, the resulting ground track of the nominal Molniya orbit
looks like in Fig.~\ref{f01}. Note that a (draconic) period of $P = 12$~hr
allows the satellite to reach its apogee in the northern hemisphere twice a day, and
$180$ deg apart in longitude. In addition to the ``Russian apogee'' (that
allowed each Molniya satellite to be on-sight from the USSR for up to 10
hours, \cite{turkiewicz75}), the supplementary North-American pass ensured, among others, a double
visibility from the US and USSR and a stable link between Russia and Cuba, 
indeed a strategic advantage along the years of the Cold War.


\subsection{Current theoretical framework}

For its special interest, the Molniya orbit has been the subject of a full range of studies
in the astrodynamical literature, starting from the '60s. 
Quite remarkably, two opposite ways to assess the dynamical problem have been pursued along the years.
In fact, the prevailing approach in most of the '60s papers is to consider the Molniya satellites
as clean gravitational probes to firmly assess the $J_2$ (and higher-order) geopotential term \cite{murphy68}, 
still poorly known at that time.

On the contrary, the focus reversed in the '70s, when the Molniya orbit itself became the subject
of investigation by inspecting the different sources of perturbation, especially dealing
with the lunisolar action as a source (together with $J_2$) of long-term effects in the evolution 
of orbital parameters and the intervening effect of the atmospheric drag to severely constrain the
satellite lifetime.

Both these aspects were first reviewed by \cite{kinghele75} and \cite{lidov76}, 
leading to estimate for the Molniya satellites an expected lifetime within 7-13 years, as a reference
figure.
The combined physical mechanism at work is correctly envisaged in the studies, with lunisolar
perturbations as a main player to act on orbital eccentricity $e$ (leaving untouched, however, the
semi-major axis $a$ and therefore the period $P$). As a consequence of a periodical change in 
$e$, the satellite perigee will decrease until possibly magnifying the effect of atmospheric drag
(especially under a favoring solar activity to ``inflate'' Earth's ionosphere).
The drag will then dissipate orbital energy such as to circularize the orbit at lower $a$ and shorter 
$P$, thus enabling the satellite's fatal re-entry in the atmosphere as a wild fireball.

The effect of the Earth's gravitational potential on the Molniya's period evolution was carefully considered in the general study presented in \cite{sochilina82}. The role of solar radiation pressure and of the Poynting-Robertson effect on high area-to-mass satellites were considered in \cite{sun13} and \cite{kuznetsov15}, respectively.

The purpose of this paper is to investigate what reasonable approximation of the models can ``reconstruct''
 the mean evolution of the orbital eccentricity given by the observational data. The approach is bottom-up, aiming at a synergy between
the observed dynamics and the mathematical modeling. The results will support a future analysis based on a dynamical systems theory approach and a practical contextualization of chaotic dynamics.

\subsection{Oblateness effect and third-body perturbation}

Since the focus is the long-term evolution of the orbital elements of the satellite, the Lagrange planetary equations \cite{Battin} are applied to various averaged perturbing contributions. 

The secular $J_{2}$ Earth's disturbing potential, $\bar{\mathcal{R}}_{J_{2}}$, is obtained by averaging over the mean anomaly $M$ (fast variable), the $J_{2}$ potential, namely,
\begin{align}
	\bar{\mathcal{R}}_{J_{2}} = \frac{1}{2\pi}\int_{0}^{2\pi}\mathcal{R}_{J_{2}} \textrm{d}M,
\end{align}
where
\begin{equation}
 \mathcal{R}_{J_{2}}=
\frac{J_{2}\mu_{\Earth} r_{\Earth}^2\big(3 \sin^{2}\phi -1\big)}{2r^3}.
\end{equation}
Here $r$ denotes the geocentric distance, $\phi$ the geocentric latitude, $r_{\Earth}$ and $\mu_{\Earth}$ the mean equatorial radius and gravitational parameter, respectively. The final expression written in terms of the orbital elements reads \cite{K66}
\begin{equation}\label{eq:Rj2}
\bar{\mathcal{R}}_{J_{2}}=
\frac{J_{2}\mu_{\Earth}r_{\Earth}^{2}}{4a^{3}(1-e^{2})^{3/2}}\big(2-3\sin^{2}i\big).
\end{equation}

Concerning the third-body disturbing potential, following \cite{celletti17}, the solar non-averaged expression can be written as
\begin{equation}\label{eq:rsun}
\begin{aligned}
\mathcal{R}_{\Sun}&=\mu_\Sun\sum_{l=2}^{\infty}\sum_{m=0}^{l}\sum_{p=0}^{l}\sum_{h=0}^{l}\sum_{q=-\infty}^{\infty}\sum_{j=-\infty}^{\infty}\frac{a^l}{a_\Sun^{l+1}}\epsilon_m\frac{(l-m)!}{(l+m)!}\mathcal{F}_{lmph}(i,i_\Sun)\times\\
&H_{lpq}(e)G_{lhj}(e_\Sun)\cos{\phi_{lmphqj}},
\end{aligned}
\end{equation}
where the geocentric orbital elements of the Sun (denoted by the astronomical symbol $\Sun$) and the spacecraft are referred to the equatorial plane, $i$ is the inclination, $\mathcal{F}_{lmph}(i,i_\Sun)$ is the product of two Kaula's inclination functions, $H_{lpq}(e)$ and $G_{lhj}(e_\Sun)$ are Hansen coefficients, and 
{\small
$$\phi_{lmphqj}=(l-2p)\omega+(l-2p+q)M-(l-2h)\omega_\Sun-(l-2h+j)M_\Sun+m\Omega,$$
}
with $\omega$ the argument of pericenter and $\Omega$ the longitude of the ascending node. For the Moon, the non-averaged disturbing potential can be written as
{\small
\begin{equation}\label{eq:rmoon}
\begin{aligned}
\mathcal{R}_{\Moon}&=\mu_\Moon
\sum_{l=2}^{\infty}
\sum_{m=0}^{l}
\sum_{p=0}^{l}
\sum_{s=0}^{l}
\sum_{q=0}^{l}
\sum_{j=-\infty}^{\infty}
\sum_{r=-\infty}^{\infty}(-1)^{m+s}(-1)^{k_1}\frac{a^l}{a_\Moon^{l+1}}
\epsilon_m\epsilon_s\frac{(l-s)!}{2(l+m)!}\mathcal{F}_{lmp}(i)\times\\
&\mathcal{F}_{lsq}(i_\Moon)H_{lpj}(e)G_{lqr}(e_\Moon)\left((-1)^{k_2}U_{lm-s}\cos{\phi^{+}_{lmpjsqr}}+(-1)^{k_3}U_{lms}\cos{\phi^{-}_{lmpjsqr}}\right),
\end{aligned}
\end{equation}
}
where the geocentric orbital elements of the Moon (denoted by the astronomical symbol $\Moon$) are referred to the ecliptic plane, and 
{\small
$$\phi^{\pm}_{lmpjsqr}=(l-2p)\omega+(l-2p+j)M+m\Omega\pm(l-2q)\omega_\Moon\pm(l-2q+r)M_\Moon \pm s(\Omega_\Moon-\pi/2)-y_s\pi.$$
}
For further details on the functions $\mathcal{F}, H,G,U$, the coefficients $\epsilon_m$, $\epsilon_s$, $k_1$, $k_2$, $k_3$ and $y_s$ and on the general expressions above, the reader can refer to \cite{celletti17}. 

The singly-averaged equations of motion (i.e., averaged over $M$), considering the second order of 
the third-body series expansion can be found in \cite{chao05}.  In particular, the approximations
\begin{align}
\left\{
\begin{aligned}
&\bar{\mathcal{R}}_{\Sun}=\frac{1}{2\pi}\int_{0}^{2\pi} \mathcal{R}_{\Sun} \, \textrm{d}M, \notag \\
&\bar{\mathcal{R}}_{\Moon}=\frac{1}{2\pi}\int_{0}^{2\pi} \mathcal{R}_{\Moon} \, \textrm{d}M,
\end{aligned}
\right.
\end{align}
are obtained by retaining the uplets satisfying the constraints $l-2p+q=0$ and $l-2p+j=0$ in the Fourier-like expansions given by Eqs.\,(\ref{eq:rsun})  and (\ref{eq:rmoon}), respectively.  

A specific analysis for HEO of singly and doubly-averaged equations of motion was given in \cite{colombo19}. 
Similarly to the singly-averaged expressions, 
the doubly-averaged disturbing functions (i.e., averaged over the mean anomaly 
$M$ and the Sun's and Moon's anomalies, $M_{\Sun}$ and $M_{\Moon}$)
\begin{align}
\left\{
\begin{aligned}
&\bar{\bar{\mathcal{R}}}_{\Sun}=\frac{1}{2\pi}\int_{0}^{2\pi} \bar{\mathcal{R}}_{\Sun} \, \textrm{d}M_{\Sun}, \notag \\
&\bar{\bar{\mathcal{R}}}_{\Moon}=\frac{1}{2\pi}\int_{0}^{2\pi} \bar{\mathcal{R}}_{\Moon} \, \textrm{d}M_{\Moon},
\end{aligned}
\right.
\end{align}
are obtained by selecting the uplets satisfying the constraints $l-2h+j=0$ and $l-2p+r=0$ in the  expansions given by Eqs.\,(\ref{eq:rsun})  and (\ref{eq:rmoon}), besides the constraints on the upsets imposed by the singly-averaged hypothesis. 

In \cite{celletti17}, the Molniya 1-81, 1-86 and 1-88 orbits (orbit \#32, 35, 36, respectively, of the convention used in the next section) were analyzed by computing the corresponding Fast Lyapunov Indicators (finite time variational chaos indicators) in the $(e,\omega)$ plane focusing on the resonance $2\dot{\omega}=0$, by assuming a second order expansion (i.e., truncating the expansions in Eqs.\,(\ref{eq:rsun}) and (\ref{eq:rmoon}) to $l=2$), averaged over the mean motion of the satellite and over the mean motion of the third body. The so-defined doubly-averaged model will be used also in the analysis of this work, but not limiting ourselves necessarily to the second order expansion.

With regard to the lunisolar perturbation on Molniya orbits, \cite{kinghele75} identified in $\Omega$ a critical parameter for the long-term evolution of the orbits. This issue was thoughtfully considered also in later studies (especially \cite{anselmo06} and 
\cite{kolyuka09}). In particular, by means of numerical investigations, considering special initial conditions for $\Omega$, these studies showed that the Molniya's lifetime could be 
extended to a very long timespan (of the order of 100-200 years). 
The importance of the role of $\Omega$ on the satellite's lifetime was remarked also in several Medium Earth Orbit studies, including Galileo's parameters \cite{AetAl16,R15,gkolias19}. 

Finally, \cite{zhu14} and \cite{zhu15} analyzed the TLE sets of the same satellites considered in this work, focusing on the dynamics associated with the semi-major axis and on the one associated with the eccentricity, respectively. That is,  they considered the effect of the tesseral harmonics on the one hand, and of the lunisolar perturbations, on the other hand. For the eccentricity, in particular, they developed a dynamical model on the basis of the harmonics $2\omega$, $2\omega+\Omega$, $-2\omega+\Omega$.

\begin{table*}[!thbp]
\centering
\caption{Initial mean elements for the Molniya satellites considered. \# identifies the sequential number of the orbit analyzed, the second, the third and fourth columns report the official identification along with the launch date, $t_0$ (MJD) the initial epoch when the orbital elements are referred to, $a$ is the semi-major axis (km), $e$ the eccentricity, $i$ the inclination (deg), $\Omega$ the longitude of the ascending node (deg), $\omega$ the argument of pericenter (deg). On the choice of these initial conditions, please refer to the text.}
\label{tab:TLE_ic}
\scriptsize
\begin{tabular}{rrrrrrrrrrr}       
\hline
\noalign{\smallskip}
\# & Molniya  & NORAD  &  Launch date& $t_0$ & $a$ & $e$ & $i$ & $\Omega\quad$ & $\omega$ & re-entry\\
   &          &  ID$\quad$ &         &    [MJD] &  [km] &     & [deg] & [deg]$\quad$ & [deg] &  (year) \\
\noalign{\smallskip}
\hline
\noalign{\smallskip}
1   & 2-09 &  7276  & April 26, 1974	& 42600.87 &  26580.72  &  0.737  &  63.40 &  310.28  &  282.57  & -   \\
2   & 2-10 &  7376  & July 23, 1974	& 44307.83 &  26574.82  &  0.722  &  64.13 &  355.47  &  274.96  & -   \\   
3   & 1-29 &  7780  & April 29, 1975	& 42559.52 &  26579.70  &  0.743  &  62.85 &  318.66  &  280.43  & -   \\
4   & 2-13 &  8015  & July 8, 1975	& 42748.56 &  26578.61  &  0.739  &  63.08 &  288.70  &  281.34  & 2018   \\
5   & 2-14 &  8195  & September 9, 1975 & 42674.15 &  26578.36  &  0.743  &  62.94 &  300.76  &  280.33  & -   \\
6   & 3-03 &  8425  & November 14, 1975 & 42746.41 &  26574.90  &  0.742  &  62.87 &  289.39  &  280.33  & 2017   \\
7   & 1-32 &  8601  & January 22, 1976  & 44306.24 &  26640.38  &  0.690  &  63.56 &  134.46  &  276.29  & -   \\
8   & 2-17 &  9829  & February 11, 1977 & 43218.70 &  26579.66  &  0.741  &  62.87 &  310.63  &  280.30  & 2020   \\
9   & 1-36 &  9880  & March 24, 1977	& 43239.65 &  26574.08  &  0.742  &  62.80 &  307.57  &  280.11  & -   \\
10  & 3-07 &  9941  & April 28, 1977	& 43283.50 &  26576.84  &  0.743  &  62.87 &  300.43  &  280.31  & 2019   \\
11  & 3-08 &  10455 & October 28, 1977  & 43476.16 &  26577.93  &  0.745  &  62.83 &  5.57    &  280.34  & -   \\
12  & 1-40 &  10925 & June 2, 1978	& 43683.63 &  26574.82  &  0.744  &  62.88 &  19.94   &  280.55  & -   \\
13  & 3-10 &  11057 & October 13, 1978  & 43806.87 &  26571.71  &  0.744  &  62.83 &  49.46   &  279.94  & -   \\
14  & 1-44 &  11474 & July 31, 1979	& 44109.27 &  26579.59  &  0.743  &  62.85 &  313.41  &  280.27  & 2017   \\
15  & 3-13 &  11896 & July 18, 1980	& 44457.07 &  26579.28  &  0.743  &  62.85 &  42.71   &  280.34  & -   \\
16  & 1-49 &  12156 & January 30, 1981  & 44728.57 &  26579.77  &  0.740  &  62.91 &  311.83  &  281.08  & -   \\
17  & 1-52 &  13012 & December 23, 1981 & 44969.94 &  26577.53  &  0.741  &  62.86 &  322.50  &  279.82  & -   \\
18  & 1-53 &  13070 & February 26, 1982 & 45043.97 &  26579.56  &  0.742  &  62.85 &  46.27   &  280.17  & -   \\
19  & 3-20 &  13875 & March 11, 1983	& 45411.79 &  26573.94  &  0.743  &  62.89 &  349.73  &  280.17  & -   \\
20  & 1-56 &  13890 & March 16, 1983	& 46357.98 &  26646.00  &  0.711  &  63.50 &  257.10  &  282.29  & -   \\
21  & 1-62 &  15214 & August 24, 1984	& 48000.94 &  26997.02  &  0.686  &  63.81 &  143.51  &  272.19  & -   \\
22  & 1-63 &  15429 & December 14, 1984 & 46065.99 &  26579.59  &  0.742  &  62.86 &  341.38  &  280.38  & -   \\
23  & 3-24 &  15738 & May 29, 1985	& 46269.33 &  26579.95  &  0.740  &  62.85 &  303.47  &  280.63  & -   \\
24  & 3-27 &  16393 & December 24, 1985 & 46443.16 &  26573.25  &  0.741  &  62.87 &  324.17  &  280.39  & -   \\
25  & 1-69 &  17078 & November 15, 1986 & 46765.04 &  26579.50  &  0.742  &  62.87 &  326.26  &  280.24  & -   \\
26  & 3-31 &  17328 & January 22, 1987  & 46861.22 &  26579.75  &  0.742  &  62.87 &  306.26  &  280.46  & -   \\
27  & 1-71 &  18946 & March 11, 1988	& 47238.66 &  26571.85  &  0.742  &  63.06 &  300.55  &  280.30  & -   \\
28  & 1-75 &  19807 & February 15, 1989 & 47619.22 &  26579.98  &  0.742  &  63.00 &  335.96  &  280.15  & -   \\
29  & 1-80 &  21118 & February 15, 1991 & 48366.38 &  26575.88  &  0.742  &  62.86 &  314.71  &  280.84  & -   \\
30  & 3-40 &  21196 & March 22, 1991	& 48494.46 &  26576.78  &  0.738  &  62.92 &  289.56  &  281.95  & -   \\
31  & 1-81 &  21426 & June 18, 1991	& 48433.04 &  26578.37  &  0.743  &  62.90 &  352.12  &  280.51  & -   \\
32  & 3-41 &  21706 & September 17, 1991& 48537.95 &  26579.98  &  0.742  &  62.85 &  243.44  &  280.31  & -   \\
33  & 3-42 &  22178 & October 14, 1992  & 48929.95 &  26579.97  &  0.742  &  62.86 &  269.96  &  280.45  & -   \\
34  & 1-86 &  22671 & May 26, 1993	& 49151.76 &  26577.28  &  0.743  &  62.83 &  241.84  &  280.42  & -   \\
35  & 1-87 &  22949 & December 22, 1993 & 49476.39 &  26575.27  &  0.741  &  62.96 &  329.29  &  281.60  & -   \\
36  & 1-88 &  23420 & December 14, 1994 & 49718.72 &  26578.62  &  0.743  &  62.80 &  245.40  &  280.37  & -   \\
37  & 3-47 &  23642 & August 9, 1995	& 49977.66 &  26579.35  &  0.742  &  62.89 &  295.99  &  280.59  & -   \\
38  & 1-90 &  24960 & September 24, 1997& 50722.56 &  26579.82  &  0.743  &  62.86 &  275.46  &  280.28  & -   \\
39  & 1-91 &  25485 & September 28, 1998& 51092.15 &  26576.12  &  0.744  &  62.86 &  311.80  &  280.45  & -   \\
40  & 3-50 &  25847 & July 8, 1999	& 51374.01 &  26575.06  &  0.742  &  62.86 &  6.20    &  280.33  & -   \\
41  & 3-51 &  26867 & July 20, 2001	& 52130.26 &  26571.97  &  0.743  &  62.86 &  248.91  &  280.46  & 2016   \\
42  & 1-93 &  28163 & February 18, 2004 & 53061.85 &  26559.84  &  0.736  &  62.86 &  202.79  &  288.20  & 2016 \\
 \noalign{\smallskip}
\hline
\end{tabular}					   
\end{table*}

\begin{figure}[htbp!]
\begin{center}
\includegraphics[width=0.327\columnwidth]{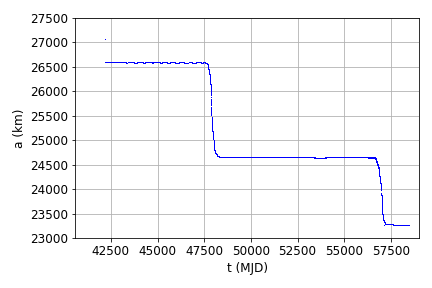}  \includegraphics[width=0.327\columnwidth]{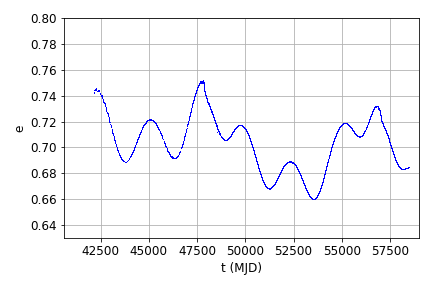} \includegraphics[width=0.317\columnwidth]{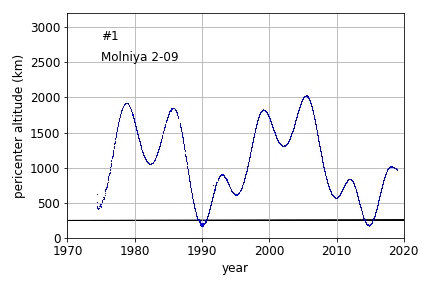}\\
\includegraphics[width=0.327\columnwidth]{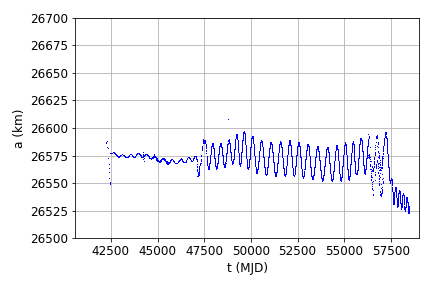}\includegraphics[width=0.327\columnwidth]{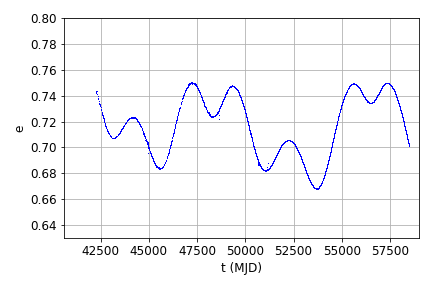} \includegraphics[width=0.317\columnwidth]{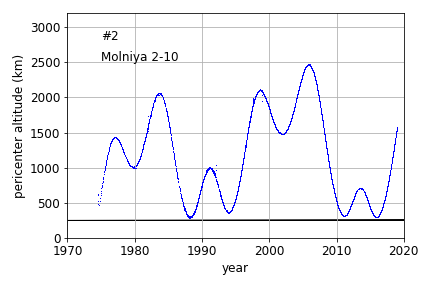}\\
\includegraphics[width=0.327\columnwidth]{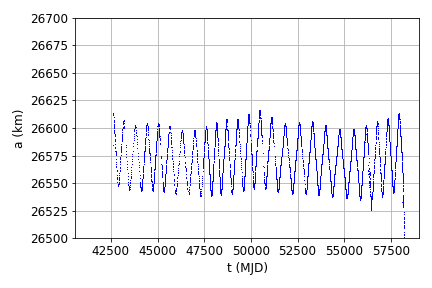}\includegraphics[width=0.327\columnwidth]{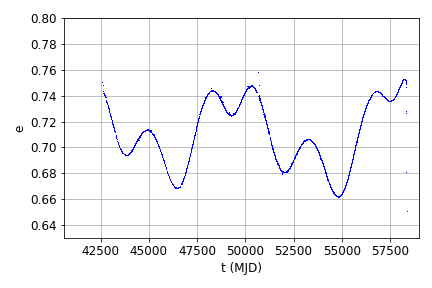} \includegraphics[width=0.317\columnwidth]{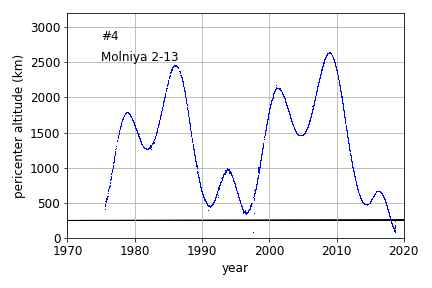}\\
 \includegraphics[width=0.327\columnwidth]{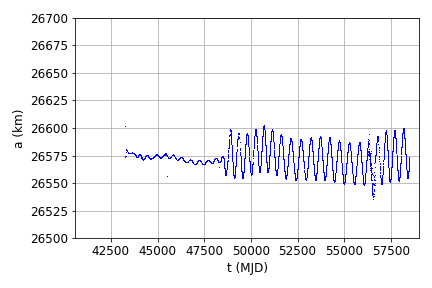} \includegraphics[width=0.327\columnwidth]{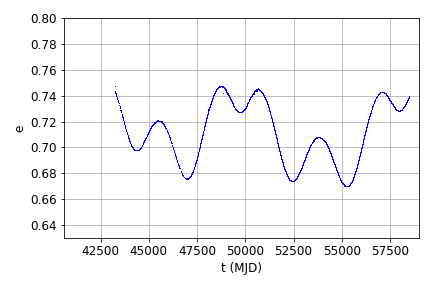} \includegraphics[width=0.317\columnwidth]{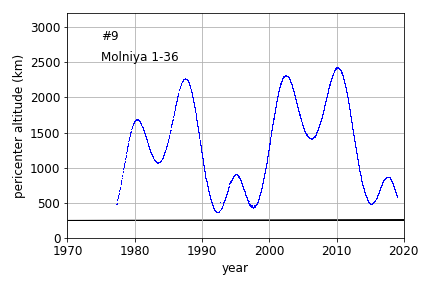}\\
  \includegraphics[width=0.327\columnwidth]{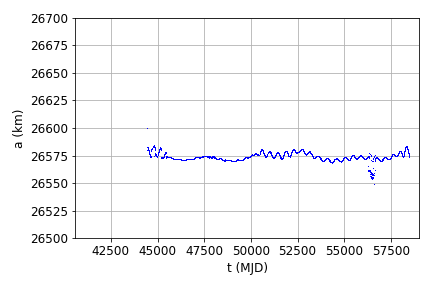} \includegraphics[width=0.327\columnwidth]{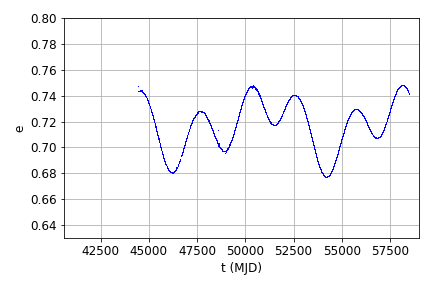} \includegraphics[width=0.317\columnwidth]{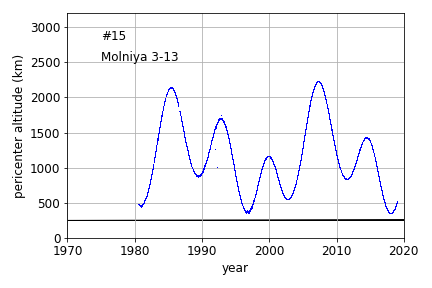}\\
   \includegraphics[width=0.327\columnwidth]{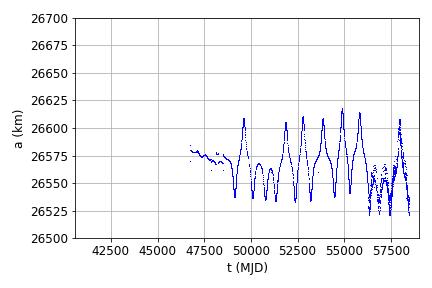} \includegraphics[width=0.327\columnwidth]{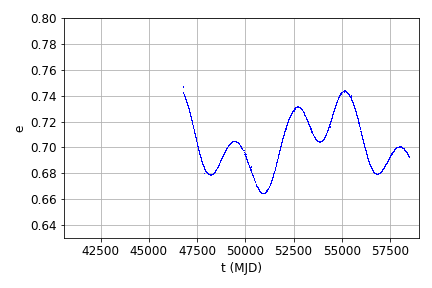} \includegraphics[width=0.317\columnwidth]{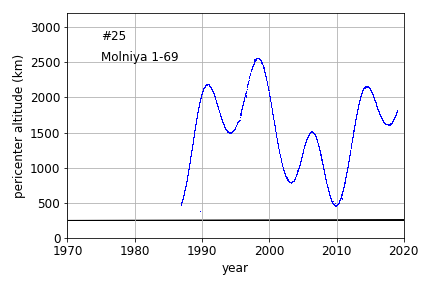}
 \caption{Semi-major axis (left; km), eccentricity (middle) and pericenter altitude (right; km) mean evolution from the TLE data of Molniya 2-09, 2-10, 2-13, 1-36, 3-13, 1-69 (\#1, 2, 4, 9, 15, 25 of Tab.~\ref{tab:TLE_ic}). On the right, the time is displayed in decimal year for the sake of clarity; also, the black horizontal line highlights 250 km of altitude.}
    \label{fig:casiscelti}
    \end{center}
\end{figure}

\subsection{Mean evolution given by TLE sets}

Table \ref{tab:TLE_ic} reports the initial conditions in mean orbital elements for the 42 HEO Molniya
satellites of the 2014 actual constellation. 
In the table, the spacecraft list is sorted in chronological
sequence, according to the launch date and NORAD identification number, as labelled.
The initial conditions displayed  correspond to the first epoch ($t_0$ reported in modified Julian Day, MJD)
where both the frozen condition $\dot\omega_{J_2}\approx 0$ and the 2:1 mean motion resonance are 
satisfied.\footnote{Except for Molniya 1-32, 1-56 and 1-62 (namely entries \#7, 20, 21 in the table)
for which the initial semi-major axis $a$ is some 1\% larger than the nominal figure.} Orbital parameters, in the table, are reported in the usual notation, being $a$ the semi-major axis (km), $e$ 
the eccentricity, $i$ the inclination (deg), $\Omega$ the longitude of the ascending node  (deg), $\omega$ the 
argument of pericenter (deg). Finally, a flag in the last column marks the seven cases of decayed satellites (with the year of the re-entry event). 

In Fig.~\ref{fig:casiscelti}, we show the time evolution of the mean semi-major axis, eccentricity and pericenter altitude obtained from the TLE sets by means of the SGP4 model \cite{NORAD} for a selected number of cases. All of them are given in the series of Figs.~\ref{fig:1serie}-\ref{fig:7serie} in Appendix A. The spurious effects, that can be noticed in the figures and that we have left for the sake of completeness, are due to the TLE. For a more complete orbital characterization, the eccentricity behavior, which is the main focus of the 
present analysis, is usefully supported with the semi-major axis evolution. 
This is done to show what kind of data we have at our disposal, and also to show when, in each case, the assumption, that we will take that the semi-major axis is constant, might fail.  

\noindent Note that the chosen initial epoch $t_0$ is usually displaced by about 1 month with respect to the launch date. The exceptions are Molniya 2-09, 2-10, 1-32, 1-56 and 1-62 (\#1, 2, 7, 20, 21) due to a non-uniform behavior in semi-major axis after the launch, that is clear from Fig.~\ref{fig:2serie} and Fig.~\ref{fig:4serie} in Appendix A for the latter three cases. The case of Molniya 2-10 will be described in detail in what follows, while for Molniya 2-09, though not visible from Fig.~\ref{fig:casiscelti}, the semi-major axis decreased almost linearly up to MJD 42500 and then started to oscillate: we have chosen to consider a good initial condition a point after this date. 

\noindent 
Moreover, we notice clear prodromic signs of an atmospheric re-entry for Molniya 2-13, 3-03, 1-44, 3-51
and 1-93 (i.e., orbits \#4, 6, 14, 41, 42); while for Molniya 2-09, 2-14, 3-10, 1-49, 1-62, 3-24, 3-27, 1-80, 3-40, 3-41, 
3-42, 1-86, 1-87, 1-88, 3-47, 1-90, 1-91 (\#1, 5, 13, 16, 21, 23, 24, 29, 30, 32-39)
a significant decrease in semi-major axis occurs, 
but the satellite remains in orbit. In all the other cases, the pericenter altitude never drops 
below $250$ km, as also noticed by \cite{zhu15}. 
As a first estimate, the data shows that the atmospheric drag can be effective to re-enter if the pericenter altitude lowers down to $210$ km.  In addition to these evident variations, we can also notice that the amplitude of oscillation in $a$ may change during the observed timespan, although in average it seems that the semi-major axis remains constant. We will see how this feature can affect the eccentricity evolution.


\section{Comparison between observational and numerical data}

We start the analysis of the astrodynamics data given by the observations, by comparing their mean evolution with the evolution that can be obtained by numerical propagation.
The initial conditions in Tab.~\ref{tab:TLE_ic} are propagated assuming the secular oblateness effect given by Eq.~(\ref{eq:Rj2}) and a doubly-averaged formulation of the lunisolar perturbation given in Eqs.~(\ref{eq:rsun})-(\ref{eq:rmoon}) under different approximations.   

In particular, following the literature mentioned before, we have tested the following physical models. 
\begin{itemize}
	\item {\it Model 1} (referred in green in the color plots): for the time evolution of the eccentricity and the inclination, the third-body perturbation is modeled using only the secular harmonics $\pm2\omega$, and the long-period ones $\pm2\omega+\Omega$ and $\Omega$, associated with the effect of both Sun and Moon; for the time evolution of $\Omega$ and $\omega$, only the oblateness effect is considered. 
This is, for $e$ and $i$, we consider
\begin{equation}\label{eq:model1}
\begin{aligned}
\frac{\textrm{d}e}{\textrm{d}t}&=-\frac{\sqrt{1-e^2}}{na^2e}{\partial\tilde{\mathcal{R}}_{3b}\over\partial \omega},\\
\frac{\textrm{d}i}{\textrm{d}t}&=-\frac{1}{na^2\sqrt{1-e^2}\sin i}\left({\partial\tilde{\mathcal{R}}_{3b}\over\partial \Omega}-\cos i {\partial\tilde{\mathcal{R}}_{3b}\over\partial \omega}\right),
\end{aligned}
\end{equation}
with
\begin{align}
\tilde{\mathcal{R}}_{3b}=\tilde{\mathcal{R}}_{\Moon}+\tilde{\mathcal{R}}_{\Sun},
\end{align}
where both $\tilde{\mathcal{R}}_{\Moon}$ and $\tilde{\mathcal{R}}_{\Moon}$ 
are obtained from Eqs.\,(\ref{eq:rsun}) and (\ref{eq:rmoon}) by taking the doubly-averaged formulation and keeping the uplets detailed below.
For the solar contribution, the set of index $(l,m,p,h,q,j)$ kept in the summation are such that\footnote{
The condition $h=1$ is derived from the fact that the doubly-averaged solar potential is independent  of the argument $\omega_{\Sun}$ for $l=2$, leading to the constraint $2-2h=0$ in Eq.\,(\ref{eq:rsun}). We refer to \cite{celletti17}, proposition $8$, for omitted details. 
}
\begin{align}
\left\{
\begin{aligned}
	& l=2, \\
	& m\in \{0,1\}, \\
	& p \in \{0,1,2\}, \\
	& h=1, \\
	& q=2p-2, \\ 
	&j=0. 
\end{aligned}
\right.
\end{align}

In  similar way, for the lunar contribution, we are led to consider the set\footnote{
Again see \cite{celletti17}, proposition $8$, for omitted details.  
}
\begin{align}
\left\{
\begin{aligned}
	& l=2, \\
	& m\in \{0,1\}, \\
	& p \in \{0,1,2\}, \\
	& s=0, \\
	& q=1, \\ 
	& j=2p-2,\\
	&r=0. 
\end{aligned}
\right.
\end{align}

For $\Omega$ and $\omega$, we assume the following rates
\begin{equation}\label{eq:J2}
\begin{aligned}
\frac{\textrm{d}\Omega}{\textrm{d}t}&=\dot\Omega_{J_2}\equiv -\frac{3}{2}\frac{J_2r^2_{\oplus}n}{a^2(1-e^2)^2}\cos i,\\
\frac{\textrm{d}\omega}{\textrm{d}t}&=\dot\omega_{J_2}\equiv \frac{3}{4}\frac{J_2r^2_{\oplus}n}{a^2(1-e^2)^2}\left(5\cos^2i-1\right).
\end{aligned}
\end{equation}
	\item {\it Model 2} (referred in red in the color plots): as in {\it model 1}, but the given lunisolar perturbations are applied also to $\Omega$ and $\omega$, namely,
\begin{eqnarray}\nonumber
\frac{\textrm{d}\Omega}{\textrm{d}t}&=&\dot\Omega_{J_2}+\frac{1}{na^2\sqrt{1-e^2}\sin i}{\partial\tilde{\mathcal{R}}_{3b}\over\partial i},\\\nonumber
\frac{\textrm{d}\omega}{\textrm{d}t}&=&\dot\omega_{J_2}+\frac{\sqrt{1-e^2}}{na^2e}{\partial\tilde{\mathcal{R}}_{3b}\over\partial e}-\frac{\cos i}{na^2\sqrt{1-e^2}\sin i}{\partial\tilde{\mathcal{R}}_{3b}\over\partial i}.\\\nonumber
\end{eqnarray}
\item {\it Model 3} (referred in black in the color plots):
the disturbing potential consists of the secular $J_{2}$ effect and the full quadrupolar ($l=2$) doubly-averaged model corresponding to both Sun and Moon.
\item {\it Model 4} (referred in yellow in the color plots): for the time evolution of the eccentricity and the inclination, we consider the doubly-averaged octupolar 
($l=3$) approximation for the Moon,  the doubly-averaged quadrupolar approximation for the Sun; for $\Omega$ and $\omega$, instead, we consider the oblateness effect and the quadrupolar doubly-averaged approximation for the third-body perturbations.
\end{itemize} 
In all the propagations, the numerical integration method is Runge-Kutta 7-8 and the ephemerides of Sun and Moon are obtained from JPL DE405 \cite{SW}. Notice that, although the purpose here is not to develop an efficient propagator for Molniya orbits, but to see what information we can extract from the available data, it is important to account for realistic lunisolar ephemerides to ensure that any discrepancies is not due to them. 

\begin{figure}[htbp!]
\begin{center}
\includegraphics[width=0.48\columnwidth]{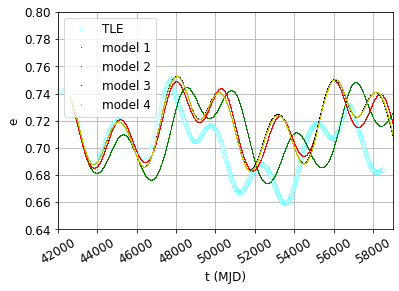}  \includegraphics[width=0.48\columnwidth]{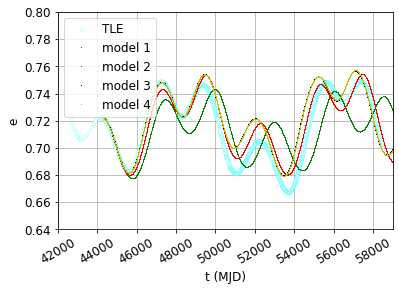} \\
\includegraphics[width=0.48\columnwidth]{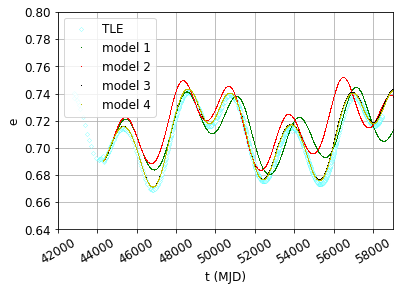}  \includegraphics[width=0.48\columnwidth]{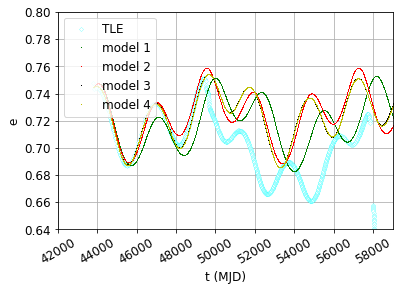}  \\ 
 \includegraphics[width=0.48\columnwidth]{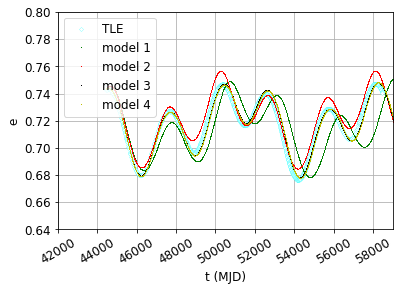} \includegraphics[width=0.48\columnwidth]{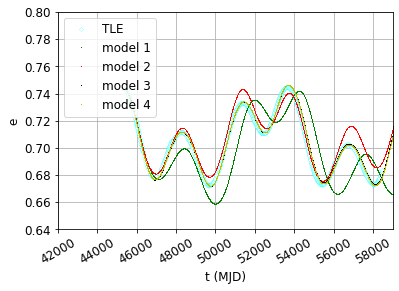} 
 \caption{Eccentricity evolution obtained by assuming different levels of third-body perturbation of $e,i,\Omega,\omega$, compared against the TLE evolution for some specific examples. More details in the text. Top: orbits  \#1 and  \#2: middle:  \#7 and  \#13; bottom:  \#15 and  \#19. All the other orbits are shown in Appendix B. }
    \label{fig:prop_semplici_examples}
    \end{center}
\end{figure}

Under all the possible hypotheses considered, the semi-major axis $a$ is constant and the problem is a two-degree-of-freedom system.
The different numerical evolutions are compared against the evolution given by the TLE (in cyan). In Fig.~\ref{fig:prop_semplici_examples}, we show some examples of the evolution of the eccentricity obtained as just described: we show three cases where we can reproduce the long-term evolution accurately and three cases where we cannot.  In Figs.~\ref{fig:prop_semplici_1serie}-\ref{fig:prop_semplici_3serie} in Appendix B, the orbital evolution of the eccentricity corresponding to all the satellites is provided.

First, we notice that almost no difference is appreciable between the results of {\it model 3} (black) and {\it model 4} (yellow). Moreover, the role of the lunisolar perturbation in  $\Omega$ and $\omega$ plays a central role in catching the real evolution of the orbit. This is appreciable comparing the cyan, green and red lines. A further improvement is obtained by adopting {\it model 3}, that can match perfectly the real evolution of the eccentricity in many cases (e.g., Molniya 1-32, 3-13 and 3-20 -- orbits \#7, 15, 19). In general, it seems sufficient to consider $l=2$, $m=0,1$, $s=0,1$ for the Moon and $l=2$, $m=0,1$ for the Sun, that is, not a full quadrupolar approximation.

\begin{figure}[ht!]
\begin{center}
\includegraphics[width=0.48\columnwidth]{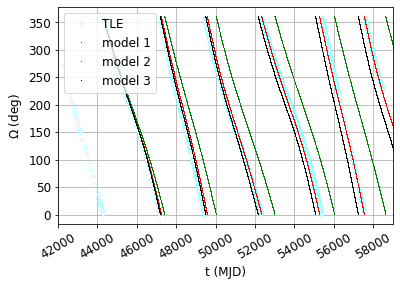}  \includegraphics[width=0.48\columnwidth]{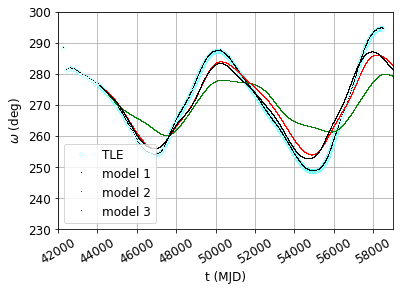} \\
\includegraphics[width=0.48\columnwidth]{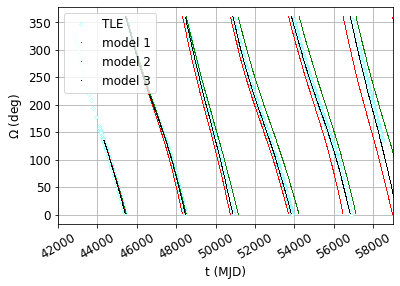}  \includegraphics[width=0.48\columnwidth]{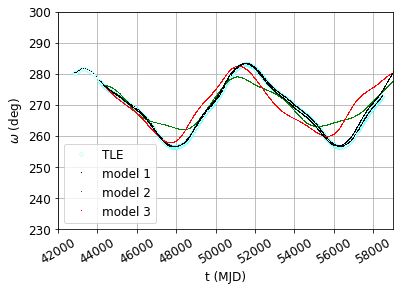}  \\ 
 \includegraphics[width=0.48\columnwidth]{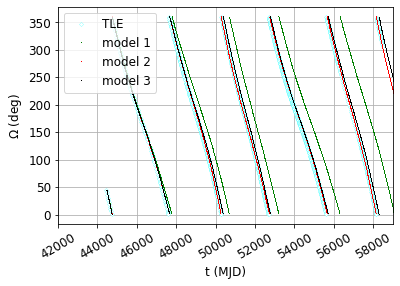} \includegraphics[width=0.48\columnwidth]{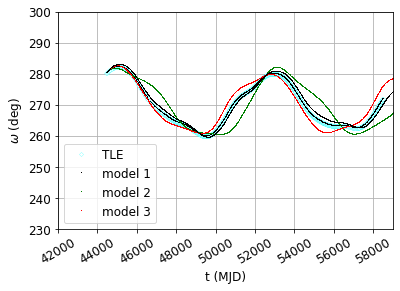} 
 \caption{Longitude of the ascending node (left) and argument of pericenter evolution (right) evolution obtained by assuming different levels of third-body perturbation of $e,i,\Omega,\omega$, compared against the TLE evolution for some specific examples. More details in the text. Top:  \#2; middle: \#7; bottom: \#15. }
    \label{fig:prop_semplici_examples_tBom_tpom}
    \end{center}
\end{figure}

\noindent
In Fig.~\ref{fig:prop_semplici_examples_tBom_tpom} the behavior of $\Omega$ and $\omega$, assuming different assumptions for the associated dynamical model, is shown for two examples.

We have tested also the following extensions to {\it model 4}:
\begin{itemize}
\item $l=3$ also for the Sun for the propagation of $e,i$;
\item $l=4$ both for Sun and Moon for the propagation of $e,i$;
\item $l=3$ for the Moon also for the propagation of $\Omega, \omega$.
\end{itemize} 
In these cases, we have not found any improvements in the qualitative behavior of the orbit (excluding the drag regime).

\begin{figure}[htbp!]
\begin{center}
\includegraphics[width=0.48\columnwidth]{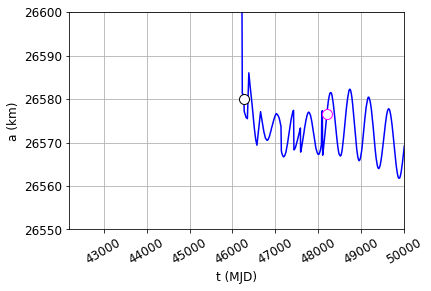}  \includegraphics[width=0.445\columnwidth]{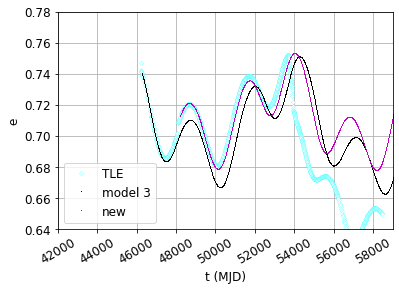} 
 \caption{On the left, a close-up view of the semi-major axis evolution given by TLE for Molniya 3-24 (orbit \#23). On the right, in black the eccentricity evolution starting before the change in the amplitude of oscillation of $a$, that can be seen at about MJD $48100$; in magenta the new orbital propagation, starting from the initial condition given in Tab.~\ref{tab:new_ic}.}
 \label{fig:case23}
    \end{center}
\end{figure}

\begin{figure}[htbp!]
\begin{center}
\includegraphics[width=0.48\columnwidth]{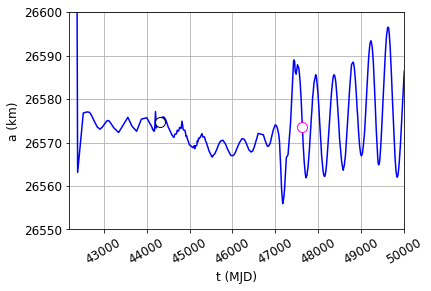}  \includegraphics[width=0.445\columnwidth]{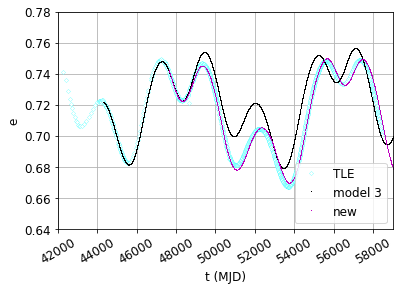} 
 \caption{On the left, a close-up view of the semi-major axis evolution given by TLE for Molniya 2-10 (orbit \#2). In black, the initial condition in semi-major axis used for the propagation shown in Fig.~\ref{fig:prop_semplici_examples}; in magenta the new initial condition whose propagation matches the behavior given by the TLE. On the right, the eccentricity evolution for the two propagations and the one given by TLE.}
 \label{fig:case2}
    \end{center}
\end{figure}

\begin{table*}[htbp!]
\centering
\caption{Initial conditions for the evolution given in Fig.~\ref{fig:case23}-\ref{fig:case2}. }
\label{tab:new_ic}
\scriptsize
\begin{tabular}{rrrrrrrrr}       
\hline
\noalign{\smallskip}
\# & Molniya  & NORAD  &   $t_0$ & $a$ & $e$ & $i$ & $\Omega\quad$ & $\omega$ \\
   &          &  ID$\quad$ &             [MJD] &  [km] &     & [deg] & [deg]$\quad$ & [deg] \\
\noalign{\smallskip}
    
2   & 2-10 &  7376  &	47628.06 	&	26573.53 	&	0.742	&	62.06	&	277.27	&	260.24	\\
23 &3-24	&	15738&	48200.85	&	26576.61&	0.712	&	63.49	&	65.71	&	278.18	\\
 \noalign{\smallskip}
\hline
\end{tabular}					   
\end{table*}

The cases that cannot be explained with {\it model 3} are Molniya 2-10, 3-10 and 3-24 (i.e.,
orbits \#2, 13, 23). 
For Molniya 3-10, in particular, this is due to the significant reduction in semi-major axis that takes place before MJD $50000$ (see Fig.~\ref{fig:3serie}). Other mismatching behaviors, but less evident, can be seen in correspondence of a significant, but less dramatic decrease in $a$, for example for Molniya 1-49 (orbit \#16 in Fig.~\ref{fig:3serie}). For Molniya 3-24 (orbit \#23), from the corresponding eccentricity evolution shown in Fig.~\ref{fig:4serie}, we can notice that the amplitude of oscillation of $a$ changes at about MJD $48100$. A new numerical propagation starting after this event has been performed assuming {\it model 3}: the magenta curve in Fig.~\ref{fig:case23} now exhibits a perfect agreement with the TLE mean evolution. Analogous considerations on the role of the initial semi-major axis can be drawn for other cases, in particular for those that show a change in the amplitude of oscillation of $a$ (orbits \#2, 5, 8, 9, 10, 11, namely Molniya 2-10,
2-14, 2-17, 1-36, 3-07 and 3-08). In Fig.~\ref{fig:case2}, we show as example the case of Molniya 2-10 (orbit \#2). The new initial conditions propagated for Molniya 3-24 and 2-10 are given in Tab.~\ref{tab:new_ic}.  Note that by assuming for Molniya 2-10 an initial epoch closer to the launch date, e.g., at MJD 42500, we obtain the same evolution depicted in black in Fig.~\ref{fig:case2}.

\noindent
Following \cite{VergerEtAl2003} (see fig. 12.7, in particular), the operational life of Molniya satellites was not longer than 6 years, that is, we cannot associate the change in amplitude to an intentional orbital maneuver. This is, however, an interesting feature that can be observed in all the cases just mentioned and that is worth to be investigated in detail in the future work focused on the semi-major axis.

Assuming that the test above ensures {\it model 3} to be reliable to predict the eccentricity evolution, barring important semi-major axis reductions, we have propagated the given initial conditions for a larger timespan -- 100 years -- to see for what cases a drop in pericenter altitude below 250 km can be attained. This is the value highlighted at the beginning for which we can observe a significant decrease in semi-major axis due to the atmospheric drag. Recall, again, that the initial conditions have been propagated considering {\it model 3}, that is, assuming $a$ constant. By excluding all the cases where it is already observed a relatively significant change in $a$, the cases for which $h_p$ lowers down to 250 km are Molniya 1-40 (orbit \#12, with possible re-entry in September 2058), Molniya 3-13 (orbit \#15, July 2039), Molniya 1-53
(orbit \#18, about April 2042),  Molniya 1-69 (orbit \#25, about April 2079) and Molniya 3-31 (orbit \#26, about August 2053). Since we are assuming a simplified model based on the oblateness effect and the lunisolar perturbations, these final dates are upper limits for the lifetime of the satellites just mentioned.

As a further consideration, the dynamics under study is known to evolve in a rather chaotic way, by which is meant that they possess the property to be sensitive to the initial condition (see, e.g., \cite{R15}). To estimate their Lyapunov times\footnote{Recall that the Lyapunov times correspond to the time needed for two nearby orbits to diverge by the Euler's number.} $\tau_{\mathcal{L}}$, 
we have relied on the variational equations derived from to the equations of motion\footnote{Thus, to estimate the Lyapunov times, we do not use the data time-series themselves by reconstructing, e.g., the phase space via Takens' delay (or also called \textit{lag}) coordinates embedding theorem \cite{fTa81}.} associated to {\it model 3}. 
We have estimated the Lyapunov times based on a 500 years numerical propagation. For all of them, 
we have found $\tau_{\mathcal{L}}$ to be roughly speaking about 10 years. Thus, the 40 years long TLE data arcs at hands represent about $4 \tau_{\mathcal{L}}$. Over this timescale, although existing, the sensitivity to the initial condition does not manifest itself so strongly as revealed by the following numerical experiment. 

\begin{figure}[htbp!]
\begin{center}
\includegraphics[width=0.49\columnwidth]{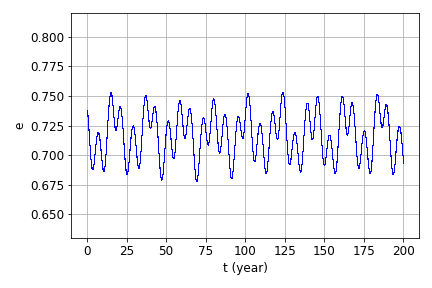}  \includegraphics[width=0.49\columnwidth]{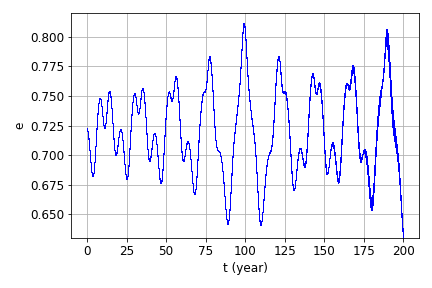} 
 \caption{Propagation of 11 neighboring initial conditions, given the same initial epoch and using {\it model 3}. The initial conditions have been taken from the propagation of the initial condition in Tab.~\ref{tab:TLE_ic} and  Tab.~\ref{tab:new_ic} for orbit \#1 (left) and  \#2 (right), respectively.}
 \label{fig:chaos}
    \end{center}
\end{figure}

\noindent
From the reconstructed nominal trajectory fitting the TLE data of the oldest survived satellites
Molniya 2-09 and 2-10 (orbit \#1 and \#2), 
we have isolated an ensemble of 
$2K=10$ initial conditions for the specific epoch $t_{0}$ by selecting the points 
$x_{n}=x(t_{n})$, $t_{n}=t_{0}+n \Delta t$, $n \in \{-K,\dots,K\}$, $\Delta t=1$ day from the nominal trajectory (i.e., neighboring points). Here $x_{n}$ denotes the Keplerian set 
$(a,e,i,\Omega,\omega)$ of geometrical elements at time $t_{n}$.
Then, we have propagated this ensemble of initial conditions forward in time over 200 years, assuming the same initial epoch (ruling the Earth-Moon configuration) for all of them and the same dynamical model. The resulting orbits, hardly distinguishable the one from the other, are shown in Fig.~\ref{fig:chaos}.
It can be noticed how the divergence among different orbits can be appreciable, but very slightly, only towards of the end of the interval of propagation of Molniya 2-10.

\begin{figure}[htbp!]
\begin{center}
  \includegraphics[width=0.32\columnwidth]{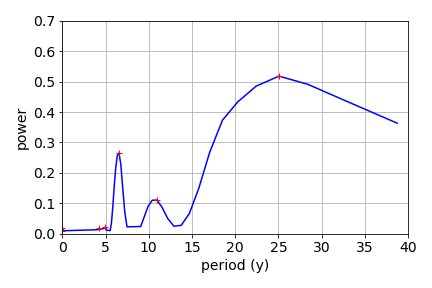}  \includegraphics[width=0.32\columnwidth]{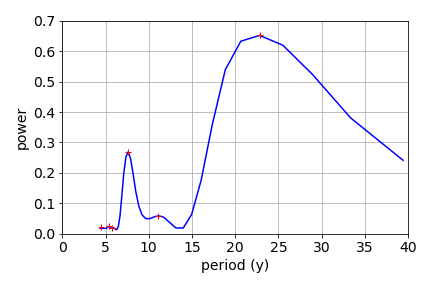}  \includegraphics[width=0.32\columnwidth]{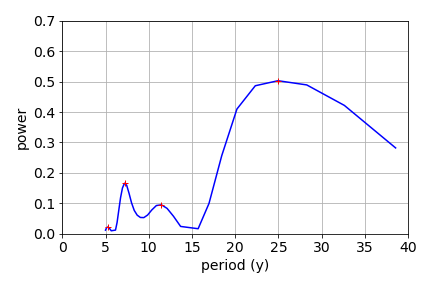}
    \caption{Example of the results obtained by applying a Lomb-Scargle procedure to the eccentricity series. From left to right: Molniya 2-09, 2-10, 1-29 (orbit \#1, 2, 3, respectively). In red, the dominant terms.}
   \label{fig:power_freq_example}
      \end{center}
\end{figure}

Finally, the eccentricity series have been processed by means of the Lomb-Scargle algorithm\footnote{LombScargle from \texttt{astropy}; given that the data are not equally spaced.} \cite{Lomb, Scargle} in order to identify the main long-term periods and compare them with the periods corresponding to the quadrupolar and octupolar doubly-averaged approximations for the third-body perturbations. Three examples are shown in Fig.~\ref{fig:power_freq_example}. The main periods detected are showed in Tab.~\ref{tab:TLE_frepeaks} for all the orbits and are related to the harmonics corresponding to $l=2$ in Eqs.~(\ref{eq:rsun})-(\ref{eq:rmoon}) (excluding the secular ones). In particular, the periods of about 7.5 years, 11 years and 24 years that stand out correspond to  $2\omega+\Omega$, $2\omega+\Omega-\Omega_\Moon$, and $2\omega+\Omega_\Moon$, respectively\footnote{Note that the corresponding information given in \cite{buzzoni19} was partially correct.}. The correspondence between the observational and the analytical approximation is obtained by assuming that the precession of $\Omega$ and $\omega$ is due to the oblateness effect and the quadrupolar doubly-averaged approximation for the third-body perturbation. The oscillations that can be noticed in the table with respect to the values just reported are due to the different initial conditions.

\begin{table*}[!t]
\centering
\caption{Main periods (years) detected by means of the Lomb-Scargle procedure for each orbit. $-$ means that the given frequency was not detected.}
\label{tab:TLE_frepeaks}
\scriptsize
\begin{tabular}{cccc}       
\hline
\noalign{\smallskip}
N & $2\dot\omega+\dot\Omega_\Moon$ & $2\dot\omega+\dot\Omega-\dot\Omega_\Moon$ &  $2\dot\omega+\dot\Omega$ \\
\noalign{\smallskip}
\hline
\noalign{\smallskip}
 1   & 25.07   & 10.93  &  6.55  \\
 2   & 22.83  & 11.12   &  7.61  \\
 3   & 24.95  & 11.46   &  7.19  \\
 4   & 22.09  & 10.76  &  8.23   \\
 5   & 24.80  & 11.39  &  7.15    \\
 6   & 24.11  & 10.51  &  8.37    \\
 7   & 25.86  & 11.76  &  7.92   \\
 8   & 23.92  & 10.99  &  7.97   \\
 9   & 27.07  & 10.98  &  7.66    \\
10  & 23.87  & 10.97  &  7.96   \\
11  & 20.75  & 10.65  &  7.44    \\
12  & 20.35  & 11.05  &  7.58    \\
13  &      -     &      -     &  7.11   \\
14  &      -     &    -       &  7.57   \\
15  & 21.35  &     -      & 7.41   \\
16  & 24.50   & 11.14  &  7.50   \\
17  & 23.99 &      -       &  7.66   \\
18  & 20.39  &     -       &  7.37  \\
19  & 22.93  &    -       &  7.64  \\
20  & 25.30  &    -       & 7.65    \\
21  & 22.41  &    -       &  7.47   \\
22  & 25.17  &   -        &   7.98    \\
23  &  -         & 10.44  &    6.87 \\
24  & 24.88  &    -       & 7.19  \\
25  & 24.67 &   11.88  &  7.82    \\
26  & 24.45  &    -      & 7.06  \\
27  & 23.44 &     -       & 7.09  \\
28  & 22.90 &  11.03   &  8.05   \\
29  & 21.29 &     -       & 7.91   \\
30  & 21.31  &   -        & 7.49  \\
31  & 20.88 &    -        & 7.33  \\
32  &       -   &   -         & 7.77  \\
33  &    -      &   -         & 7.06  \\
34  &    -      &    -       &  8.21   \\
35  &    -      &   -       &  7.95    \\
36  &    -      &   -        & 7.24 \\
37  &    -      &   -       &  7.49    \\
38  &    -     &  11.14   &  5.72 \\
39  &    -     &    -      &  6.48 \\
40  &    -     &    -      &  8.36    \\
41  &    -     &    -      &   7.28  \\
42  &   -      &    -      &   7.14 \\
 \noalign{\smallskip}
\hline
\end{tabular}					   
\end{table*}


\section{Conclusions and future directions}

In this work, we have analyzed the long-term evolution of the mean eccentricity obtained from TLE sets of the Molniya historical constellation. The analysis has considered the third-body effect as the major perturbation on the orbital eccentricity. Different assumptions on a two-degree-of-freedom dynamical model accounting for the lunisolar perturbations coupled with the oblateness effect have been compared against the observational data. The outcome shows that a quadrupolar doubly-averaged formulation represents a reliable model to depict a realistic evolution. Also, it has emerged the importance of the role of the lunisolar perturbation in the time evolution of $\Omega$ and  $\omega$. 

The role of the initial conditions turns out to be important for what concerns the semi-major axis, not only in case of a significant reduction (as expected), but also when the amplitude of oscillation varies. The corresponding behavior and the reason of the change exhibited by several satellites will be faced in a future work. In the same work, the role of tesseral harmonics and of the atmospheric drag, along with a detailed analysis of the solar activity, will be presented. 

Preliminary numerical experiments have been performed to understand whether chaotic phenomena manifest in the timespan considered. 
The conclusion is that in the considered time interval the chaotic nature of the problem does not apparently affect so strongly the dynamics. 

Finally, the work \cite{Tiziana}, just concluded,  has analyzed amplitudes and periods of the lunisolar doubly-averaged expansions up to the octupolar approximation, with the purpose of identifying the major contributions for a proper phase space description. Such Hamiltonian description has been supported by the numerical comparison provided here and will be published in a separate work.  From that theoretical description and the phase space analysis, the role of the initial $\Omega$ pointed out in the past will be clarified and a more detailed analysis on the role of chaos will be carried out.

\section*{Acknowledgements}
E.M.A. and G.T. are grateful to Tiziana Talu for the work she has carried out for her Master's thesis, that has supported the present analysis. J.D. acknowledges the financial support from naXys Research Institute.

\newpage
\section*{Appendix A}

In the following figures, we show the semi-major axis (left; km), eccentricity (middle) and pericenter altitude (right; km) mean evolution from the TLE data of the satellites reported in Tab.~\ref{tab:TLE_ic}. On the right, the time is displayed in decimal year for the sake of clarity. The black horizontal line on the right plot highlights 250 km of altitude.

\begin{figure}[htbp!]
\begin{center}
\includegraphics[width=0.327\columnwidth]{ta_zoom_new_1.png}  \includegraphics[width=0.327\columnwidth]{te_new_1.png} \includegraphics[width=0.317\columnwidth]{thp_year_1.png}\\
\includegraphics[width=0.327\columnwidth]{ta_zoom_new_2.png}\includegraphics[width=0.327\columnwidth]{te_new_2.png} \includegraphics[width=0.317\columnwidth]{thp_year_2.png}\\
 \includegraphics[width=0.327\columnwidth]{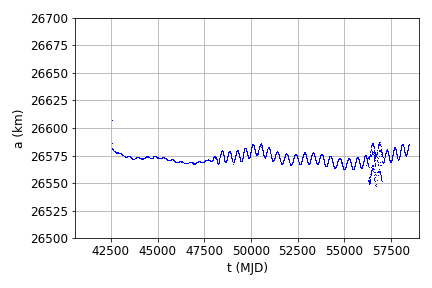}\includegraphics[width=0.327\columnwidth]{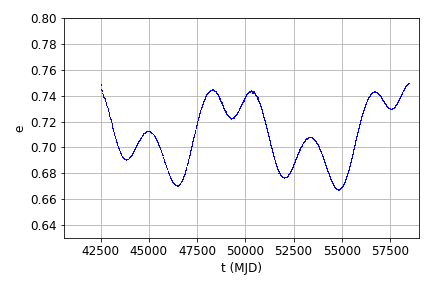} \includegraphics[width=0.317\columnwidth]{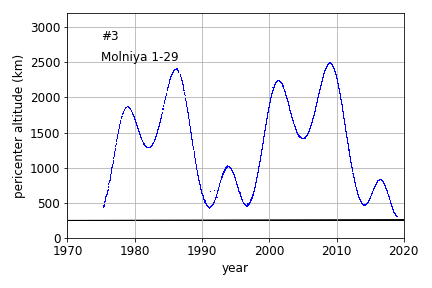}\\
\includegraphics[width=0.327\columnwidth]{ta_zoom_new_5.png}\includegraphics[width=0.327\columnwidth]{te_new_5.png} \includegraphics[width=0.317\columnwidth]{thp_year_5.png}\\
\includegraphics[width=0.327\columnwidth]{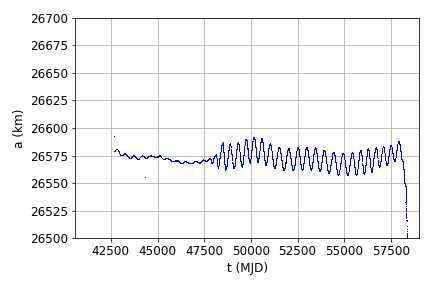} \includegraphics[width=0.327\columnwidth]{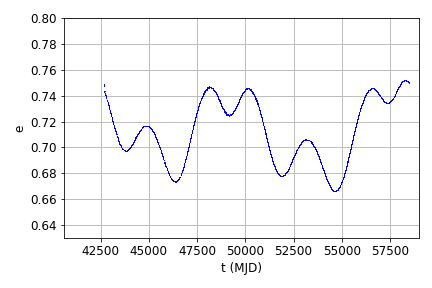} \includegraphics[width=0.317\columnwidth]{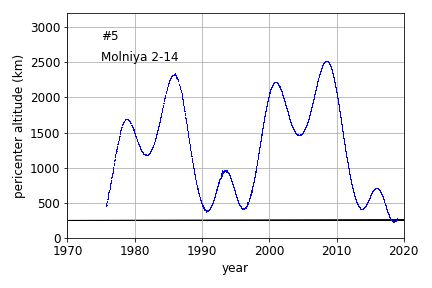}\\
 \includegraphics[width=0.327\columnwidth]{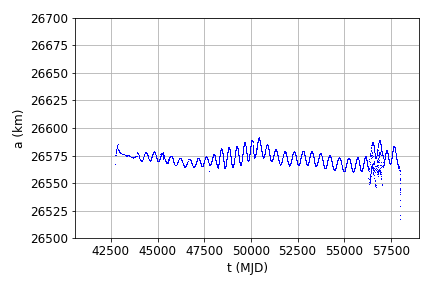} \includegraphics[width=0.327\columnwidth]{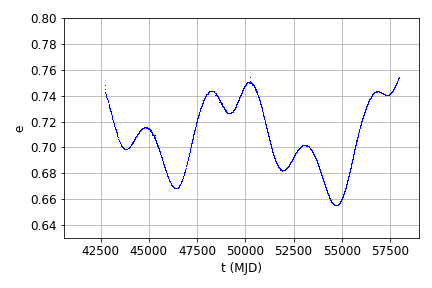} \includegraphics[width=0.317\columnwidth]{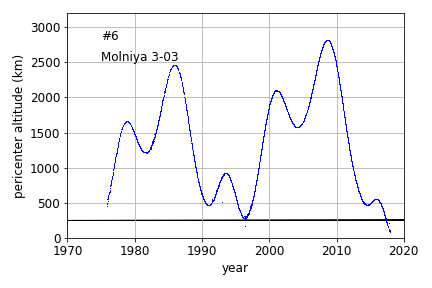}
\caption{} 
    \label{fig:1serie}
    \end{center}
\end{figure}

\noindent
In Fig.~\ref{fig:1serie}, orbits \#1-6 are reported (top to bottom).

\begin{figure}[htbp!]
\begin{center}
\includegraphics[width=0.327\columnwidth]{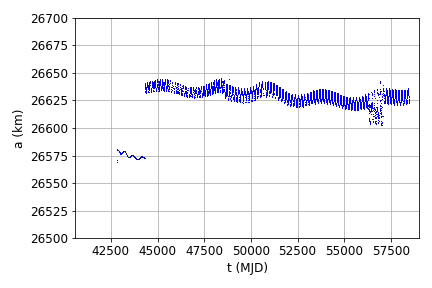} \includegraphics[width=0.327\columnwidth]{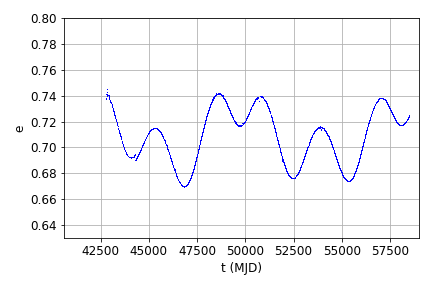} \includegraphics[width=0.317\columnwidth]{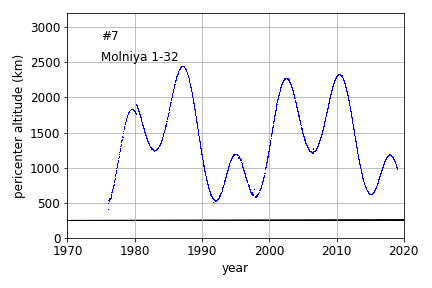}\\
\includegraphics[width=0.327\columnwidth]{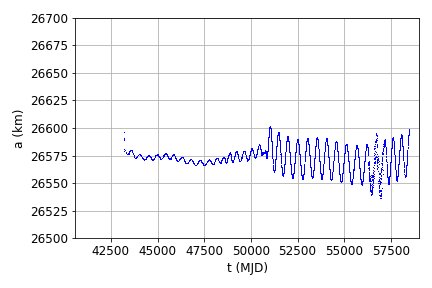} \includegraphics[width=0.327\columnwidth]{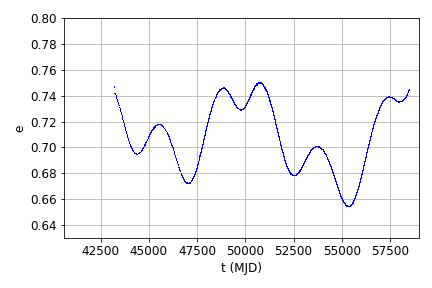} \includegraphics[width=0.317\columnwidth]{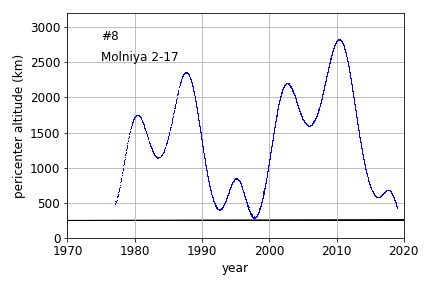}\\
 \includegraphics[width=0.327\columnwidth]{ta_zoom_new_10.png} \includegraphics[width=0.327\columnwidth]{te_new_10.png} \includegraphics[width=0.317\columnwidth]{thp_year_10.png}\\
 \includegraphics[width=0.327\columnwidth]{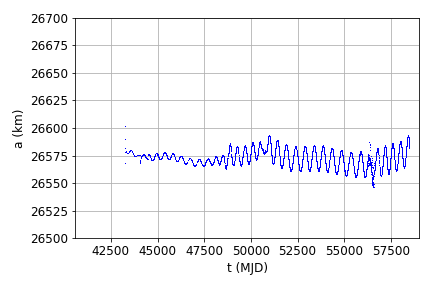} \includegraphics[width=0.327\columnwidth]{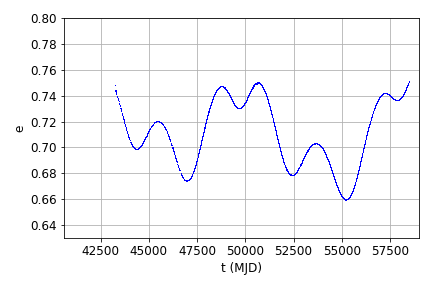} \includegraphics[width=0.317\columnwidth]{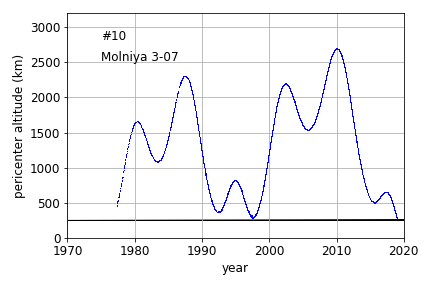}\\
  \includegraphics[width=0.327\columnwidth]{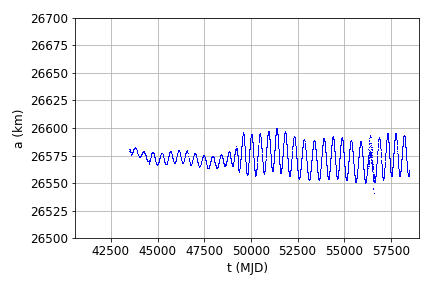} \includegraphics[width=0.327\columnwidth]{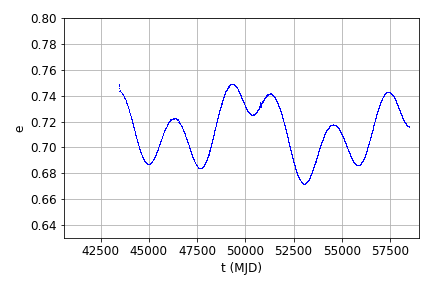} \includegraphics[width=0.317\columnwidth]{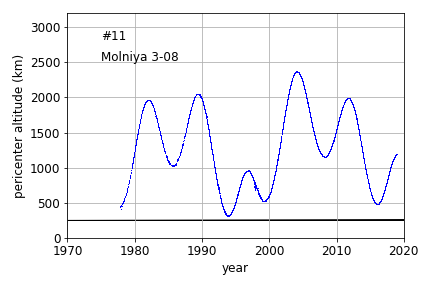}\\
 \includegraphics[width=0.327\columnwidth]{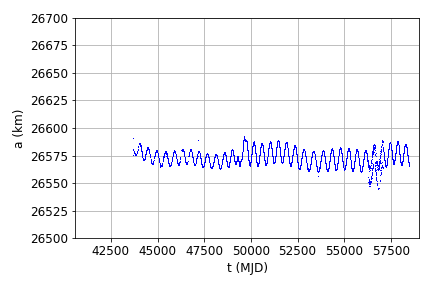} \includegraphics[width=0.327\columnwidth]{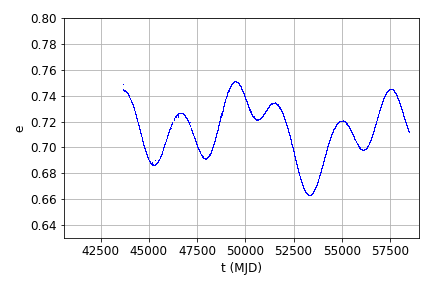} \includegraphics[width=0.317\columnwidth]{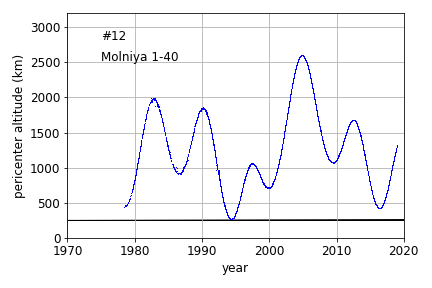}
 \caption{}
    \label{fig:2serie}
    \end{center}
\end{figure}

\noindent
In Fig.~\ref{fig:2serie}, torbits \#7-12 are reported (top to bottom).

\begin{figure}[htbp!]
\begin{center}
 \includegraphics[width=0.327\columnwidth]{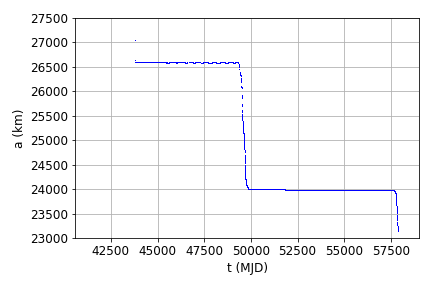}  \includegraphics[width=0.327\columnwidth]{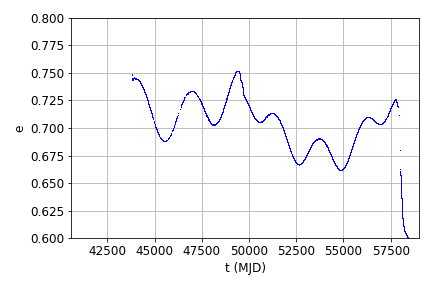} \includegraphics[width=0.317\columnwidth]{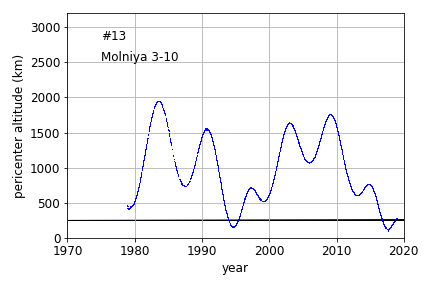}\\
\includegraphics[width=0.327\columnwidth]{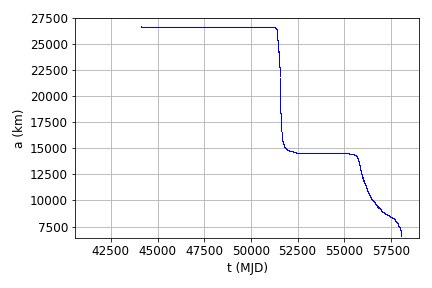} \includegraphics[width=0.327\columnwidth]{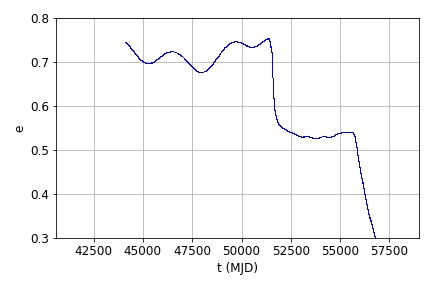} \includegraphics[width=0.317\columnwidth]{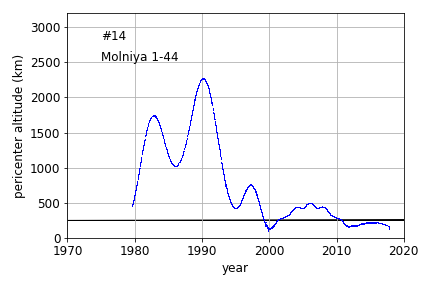}\\
 \includegraphics[width=0.327\columnwidth]{ta_zoom_new_16.png} \includegraphics[width=0.327\columnwidth]{te_new_16.png} \includegraphics[width=0.317\columnwidth]{thp_year_16.png}\\
  \includegraphics[width=0.327\columnwidth]{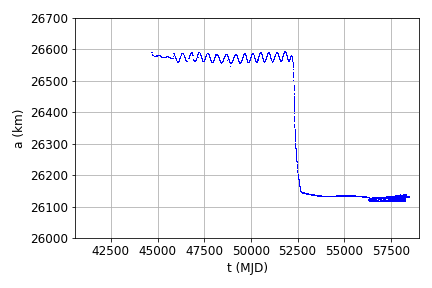}  \includegraphics[width=0.327\columnwidth]{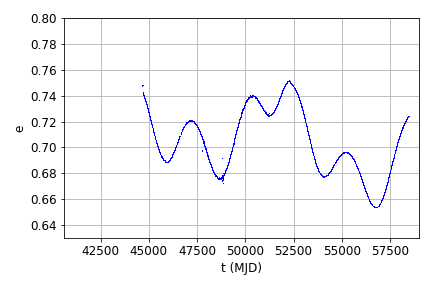} \includegraphics[width=0.317\columnwidth]{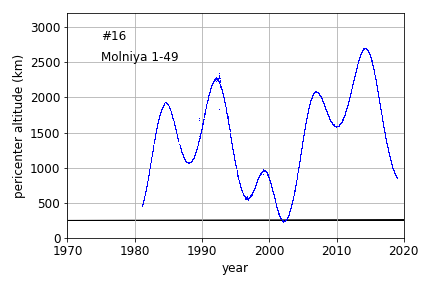}\\
 \includegraphics[width=0.327\columnwidth]{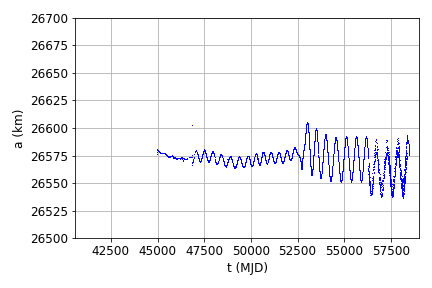} \includegraphics[width=0.327\columnwidth]{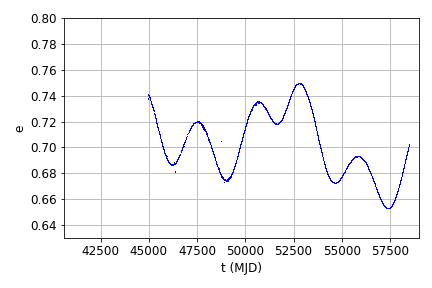} \includegraphics[width=0.317\columnwidth]{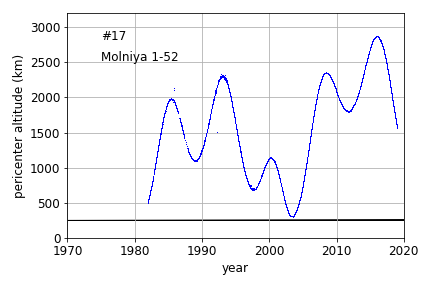}\\
 \includegraphics[width=0.327\columnwidth]{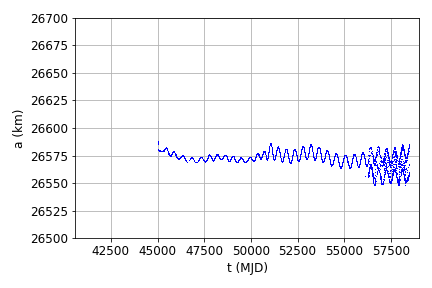} \includegraphics[width=0.327\columnwidth]{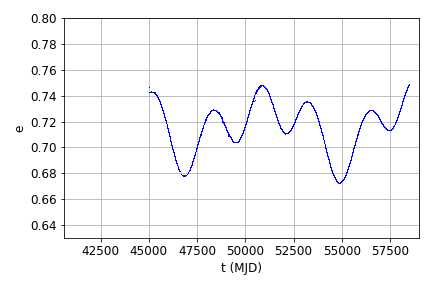} \includegraphics[width=0.317\columnwidth]{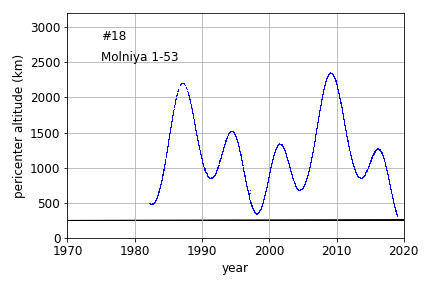}
\caption{} 
    \label{fig:3serie}
    \end{center}
\end{figure}

 \noindent
In Fig.~\ref{fig:3serie}, orbits \#13-18 are reported (top to bottom).

\begin{figure}[htbp!]
\begin{center}
 \includegraphics[width=0.327\columnwidth]{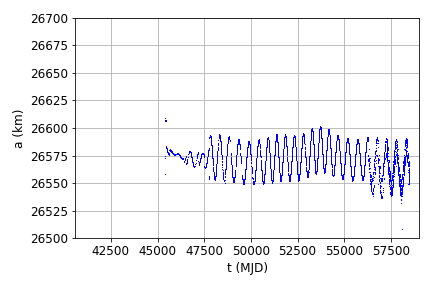} \includegraphics[width=0.327\columnwidth]{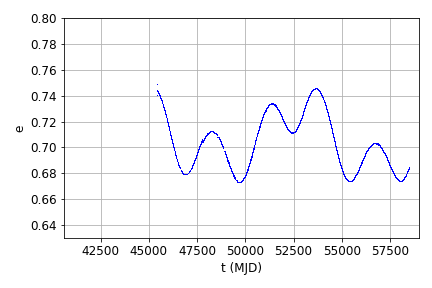} \includegraphics[width=0.317\columnwidth]{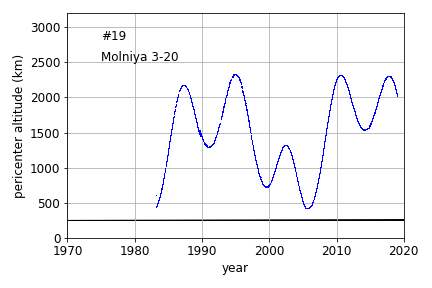}\\
\includegraphics[width=0.327\columnwidth]{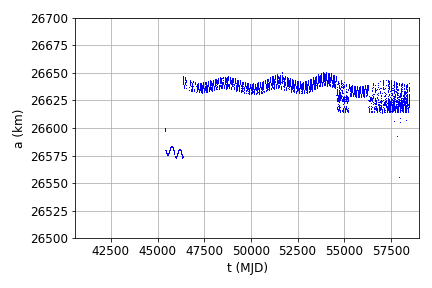} \includegraphics[width=0.327\columnwidth]{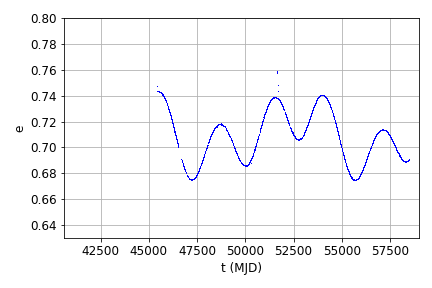} \includegraphics[width=0.317\columnwidth]{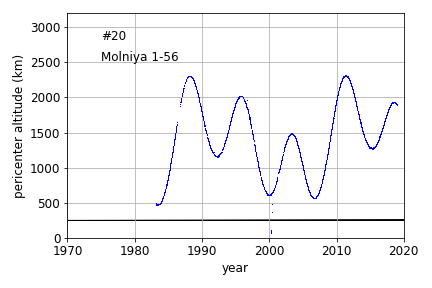}\\
\includegraphics[width=0.327\columnwidth]{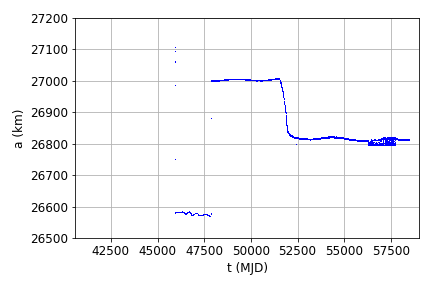}  \includegraphics[width=0.327\columnwidth]{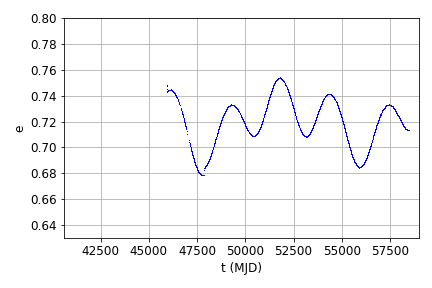} \includegraphics[width=0.317\columnwidth]{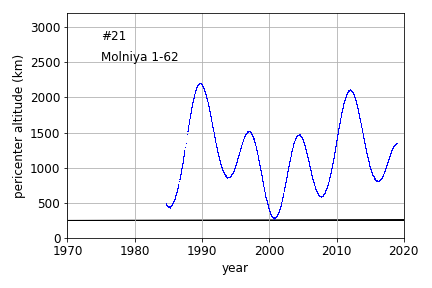}\\
 \includegraphics[width=0.327\columnwidth]{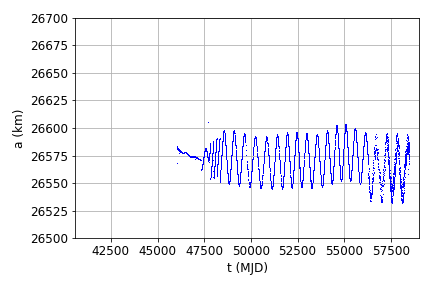} \includegraphics[width=0.327\columnwidth]{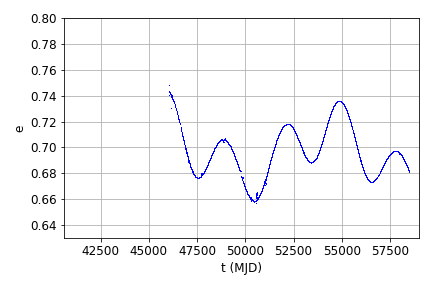} \includegraphics[width=0.317\columnwidth]{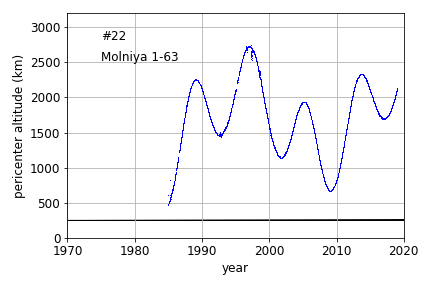}\\
 \includegraphics[width=0.327\columnwidth]{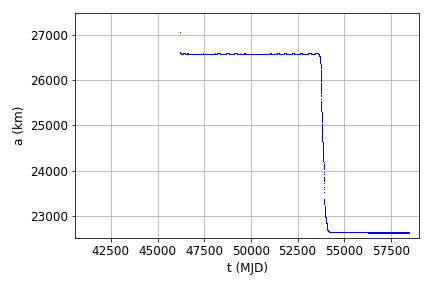} \includegraphics[width=0.327\columnwidth]{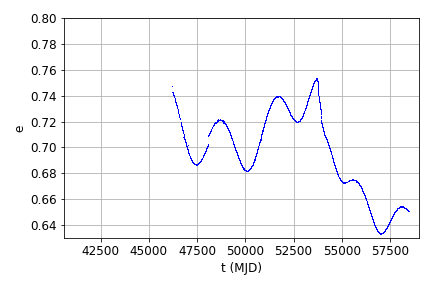} \includegraphics[width=0.317\columnwidth]{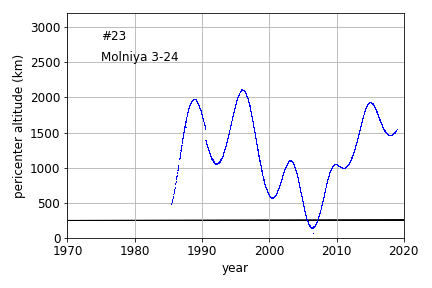}\\
\includegraphics[width=0.327\columnwidth]{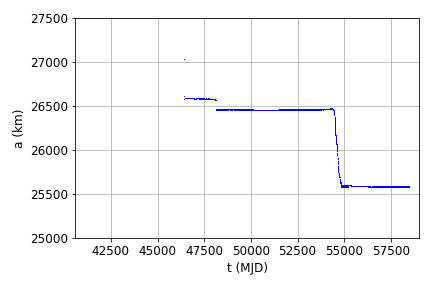} \includegraphics[width=0.327\columnwidth]{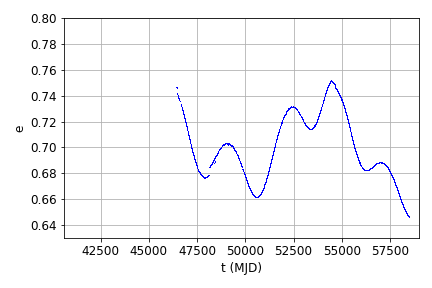} \includegraphics[width=0.317\columnwidth]{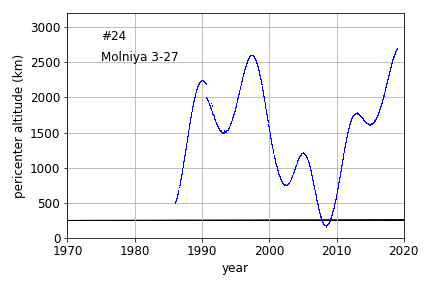}
\caption{} 
    \label{fig:4serie}
    \end{center}
\end{figure}

 \noindent
In Fig.~\ref{fig:4serie}, orbits \#19-24 are reported (top to bottom).

\begin{figure}[htbp!]
\begin{center}
 \includegraphics[width=0.327\columnwidth]{ta_zoom_new_26.png} \includegraphics[width=0.327\columnwidth]{te_new_26.png} \includegraphics[width=0.317\columnwidth]{thp_year_26.png}\\
  \includegraphics[width=0.327\columnwidth]{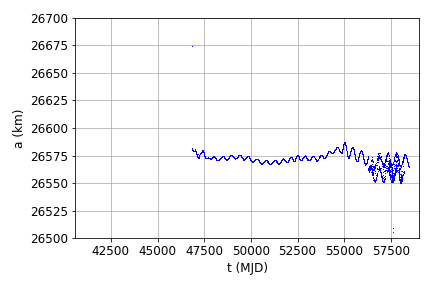} \includegraphics[width=0.327\columnwidth]{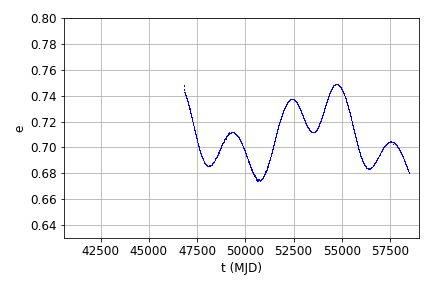} \includegraphics[width=0.317\columnwidth]{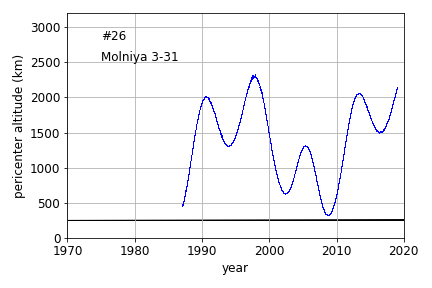}\\
 \includegraphics[width=0.327\columnwidth]{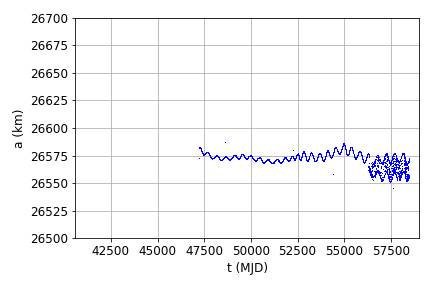} \includegraphics[width=0.327\columnwidth]{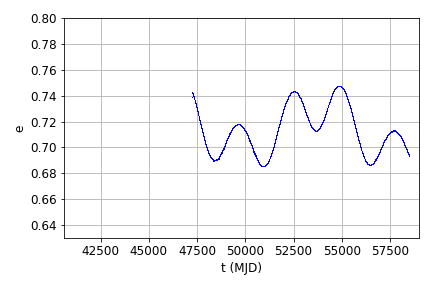} \includegraphics[width=0.317\columnwidth]{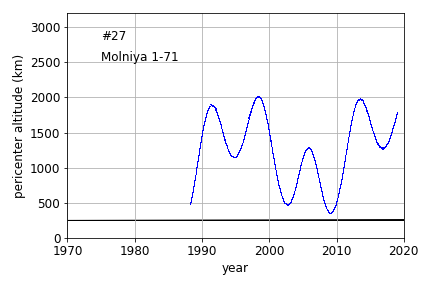}\\
 \includegraphics[width=0.327\columnwidth]{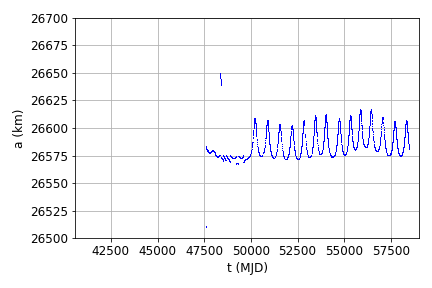} \includegraphics[width=0.327\columnwidth]{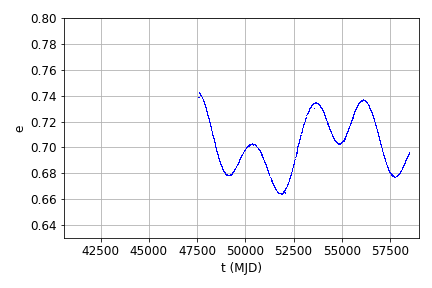} \includegraphics[width=0.317\columnwidth]{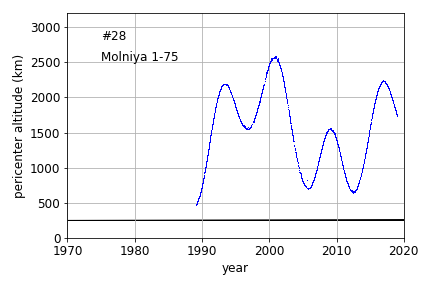}\\
 \includegraphics[width=0.327\columnwidth]{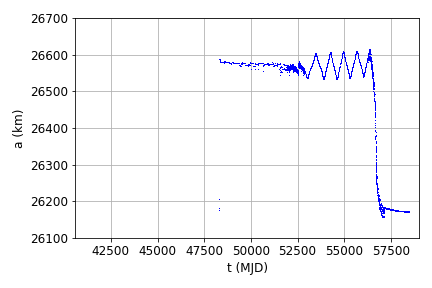}  \includegraphics[width=0.327\columnwidth]{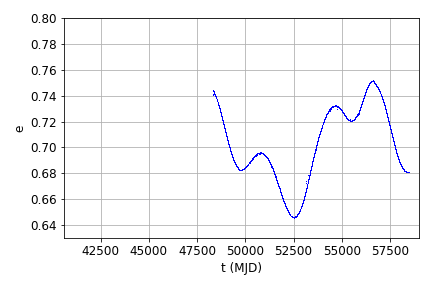} \includegraphics[width=0.317\columnwidth]{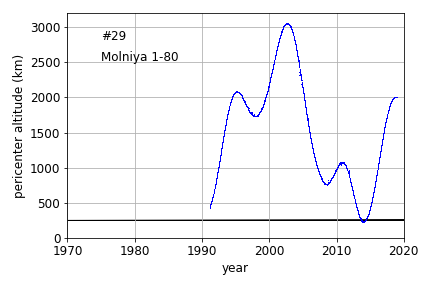}\\
\includegraphics[width=0.327\columnwidth]{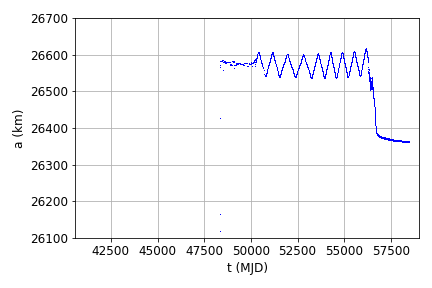}  \includegraphics[width=0.327\columnwidth]{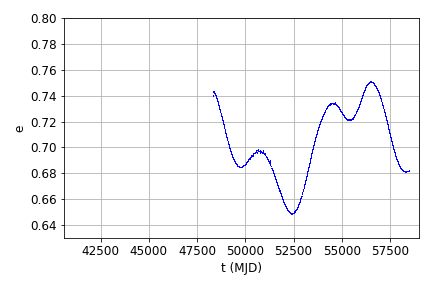} \includegraphics[width=0.317\columnwidth]{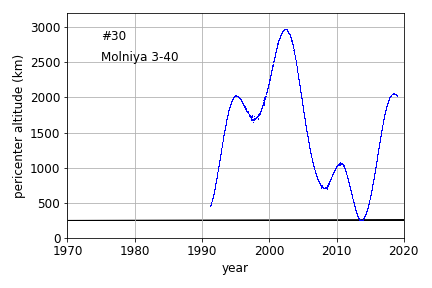}
\caption{}
    \label{fig:5serie}
    \end{center}
\end{figure}

 \noindent
In Fig.~\ref{fig:5serie}, orbits \#25-30 are reported (top to bottom).

\begin{figure}[htbp!]
\begin{center}
 \includegraphics[width=0.327\columnwidth]{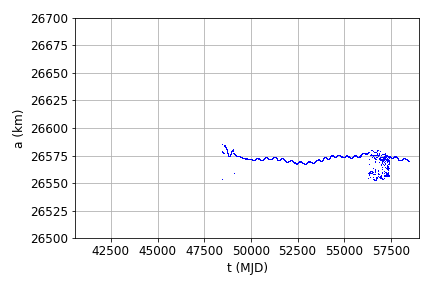} \includegraphics[width=0.327\columnwidth]{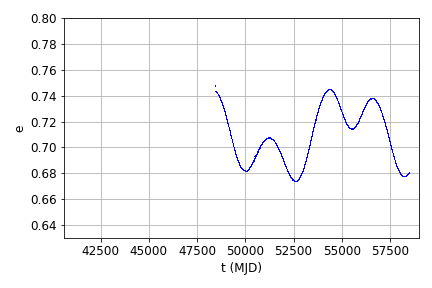} \includegraphics[width=0.317\columnwidth]{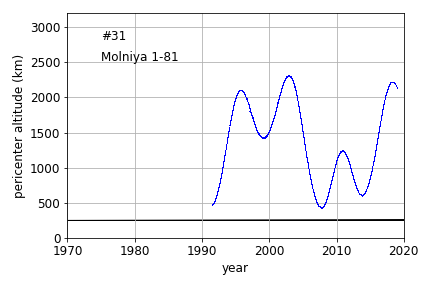}\\
\includegraphics[width=0.327\columnwidth]{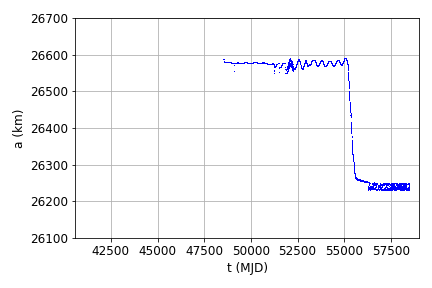}  \includegraphics[width=0.327\columnwidth]{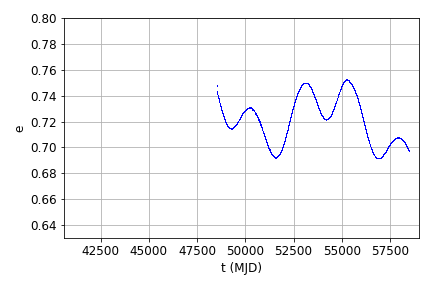} \includegraphics[width=0.317\columnwidth]{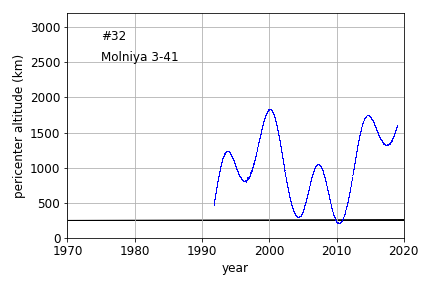}\\
\includegraphics[width=0.327\columnwidth]{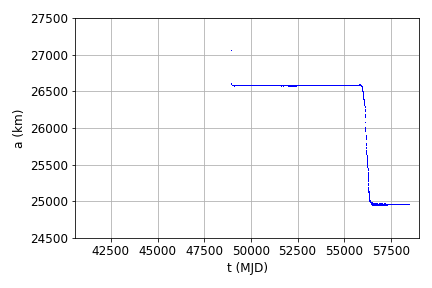} \includegraphics[width=0.327\columnwidth]{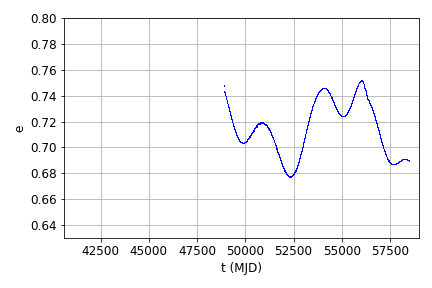} \includegraphics[width=0.317\columnwidth]{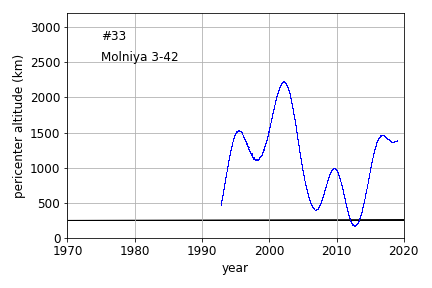}\\
\includegraphics[width=0.327\columnwidth]{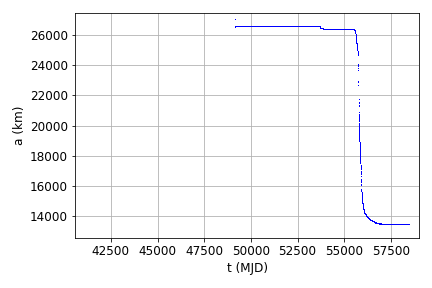} \includegraphics[width=0.327\columnwidth]{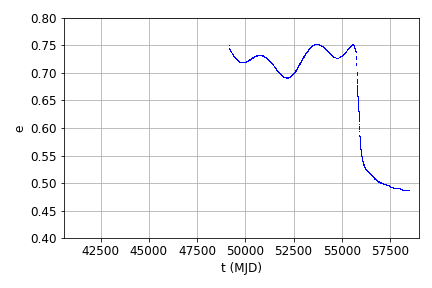} \includegraphics[width=0.317\columnwidth]{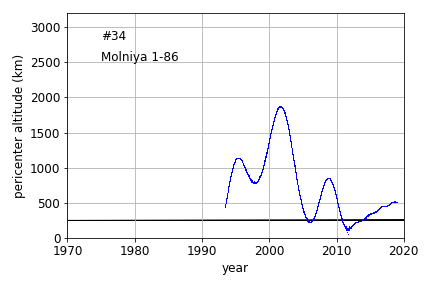}\\
\includegraphics[width=0.327\columnwidth]{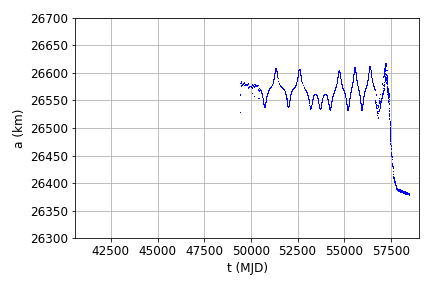} \includegraphics[width=0.327\columnwidth]{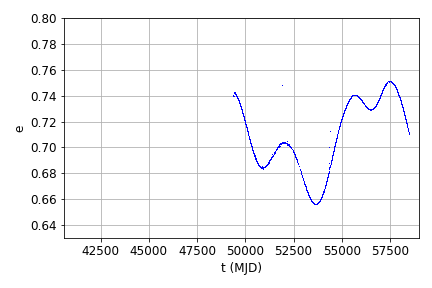} \includegraphics[width=0.317\columnwidth]{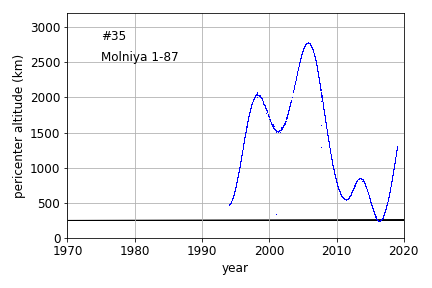}\\
\includegraphics[width=0.327\columnwidth]{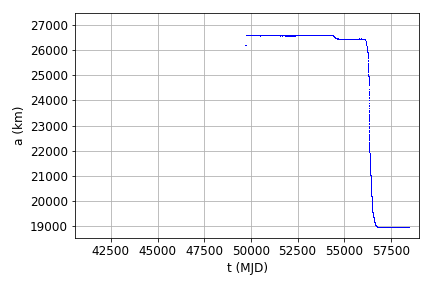} \includegraphics[width=0.327\columnwidth]{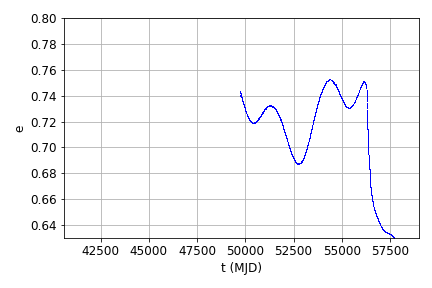} \includegraphics[width=0.317\columnwidth]{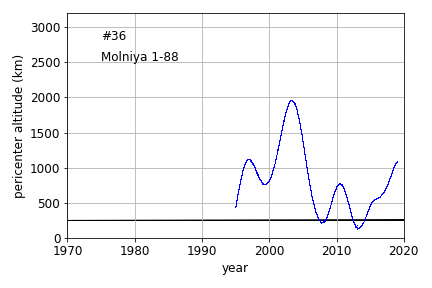}
\caption{} 
    \label{fig:6serie}
    \end{center}
\end{figure}

 \noindent
In Fig.~\ref{fig:6serie}, orbits \#31-36 are reported (top to bottom).

\begin{figure}[htbp!]
\begin{center}
\includegraphics[width=0.327\columnwidth]{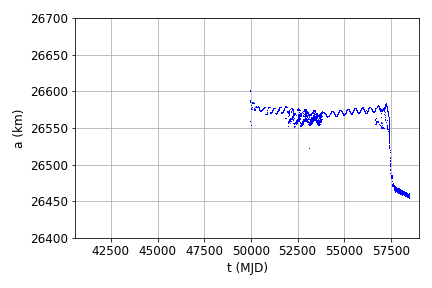} \includegraphics[width=0.327\columnwidth]{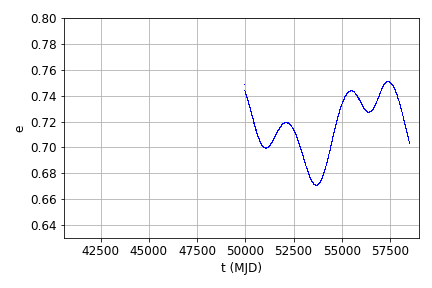} \includegraphics[width=0.317\columnwidth]{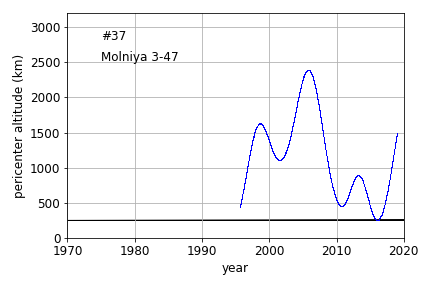}\\
\includegraphics[width=0.327\columnwidth]{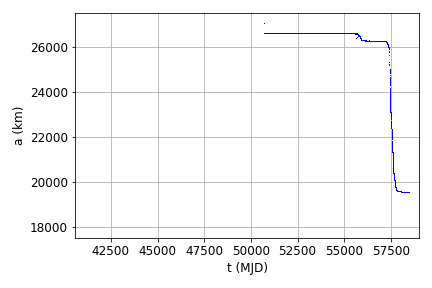}  \includegraphics[width=0.327\columnwidth]{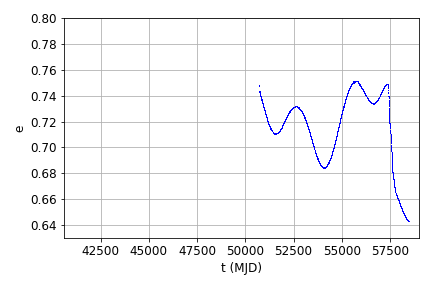} \includegraphics[width=0.317\columnwidth]{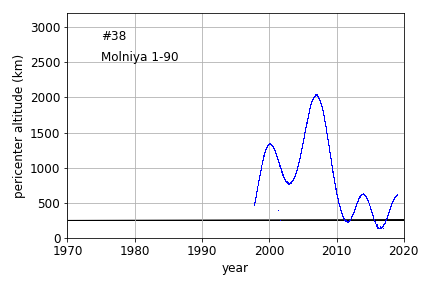}\\
\includegraphics[width=0.327\columnwidth]{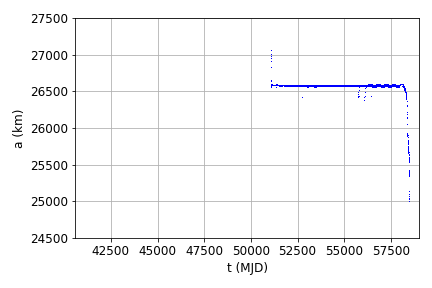}  \includegraphics[width=0.327\columnwidth]{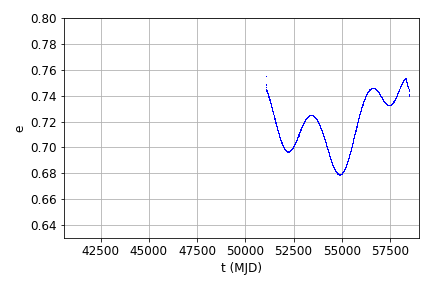} \includegraphics[width=0.317\columnwidth]{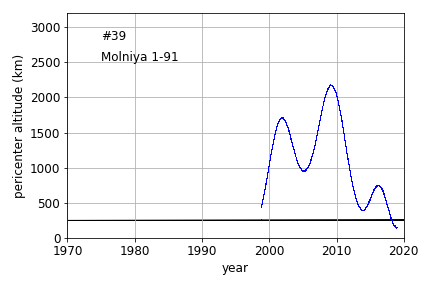}\\
 \includegraphics[width=0.327\columnwidth]{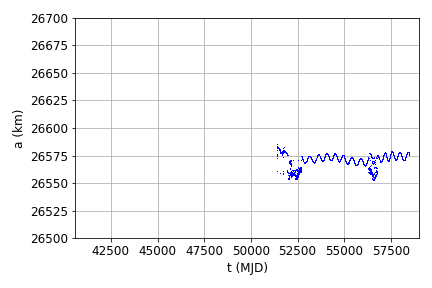} \includegraphics[width=0.327\columnwidth]{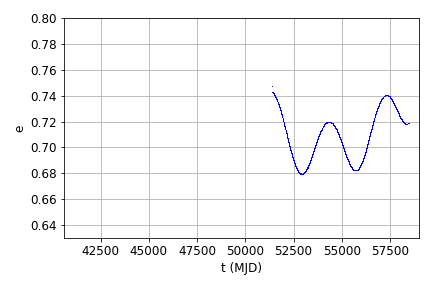} \includegraphics[width=0.317\columnwidth]{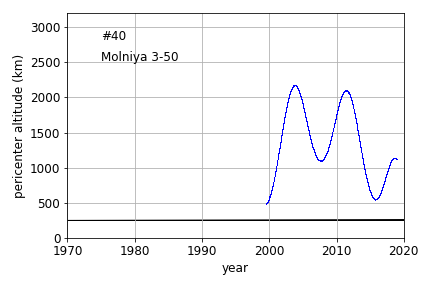}\\
\includegraphics[width=0.327\columnwidth]{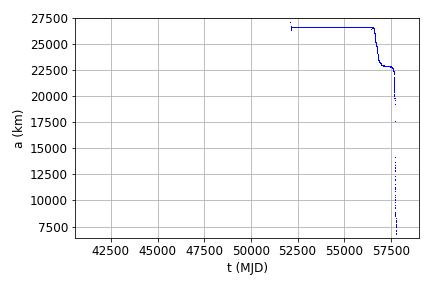} \includegraphics[width=0.327\columnwidth]{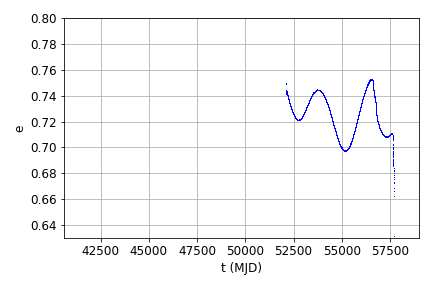} \includegraphics[width=0.317\columnwidth]{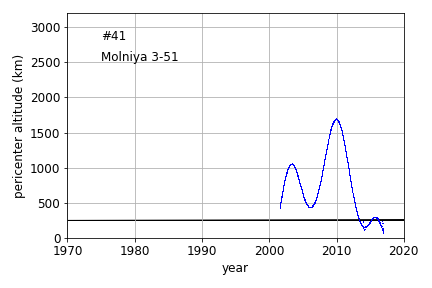}\\
\includegraphics[width=0.327\columnwidth]{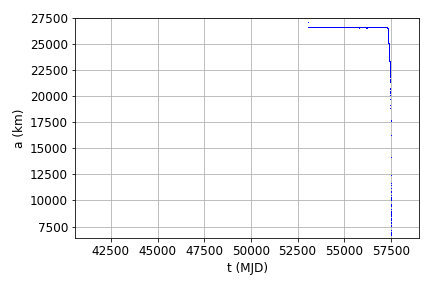} \includegraphics[width=0.327\columnwidth]{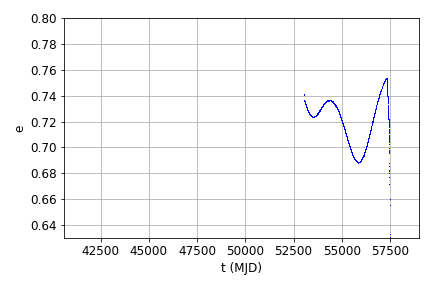} \includegraphics[width=0.317\columnwidth]{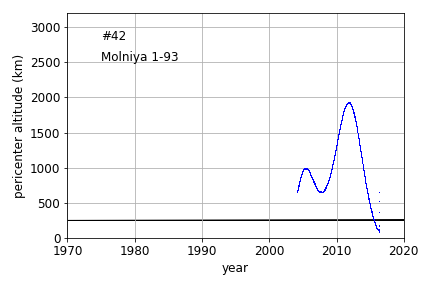}
\caption{} 
    \label{fig:7serie} 
    \end{center}
\end{figure}

 \noindent
In Fig.~\ref{fig:7serie}, orbits \#37-42 are reported (top to bottom).

\newpage
\section*{Appendix B}

In the following figures, we show the evolution obtained by assuming different levels of third-body perturbation of $e,i,\Omega,\omega$, compared against the TLE evolution. Each plot corresponds to a satellite of  Tab.~\ref{tab:TLE_ic}. The ordering is left to right, top to bottom. 

\begin{figure}[h!]
\begin{center}
\includegraphics[width=0.32\columnwidth]{te_tepropsempl_final_new_1.png}  \includegraphics[width=0.32\columnwidth]{te_tepropsempl_final_new_2.png} \includegraphics[width=0.32\columnwidth]{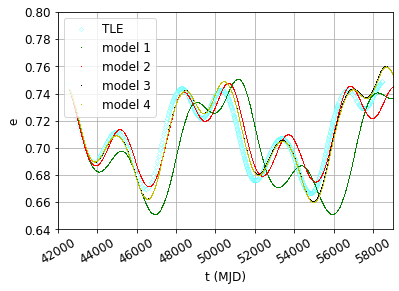}\\
\includegraphics[width=0.32\columnwidth]{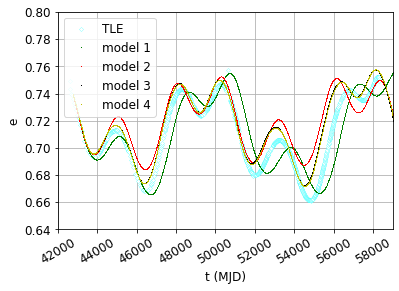}  \includegraphics[width=0.32\columnwidth]{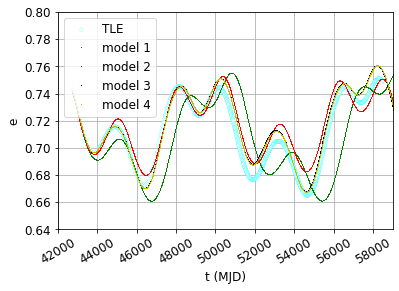} \includegraphics[width=0.32\columnwidth]{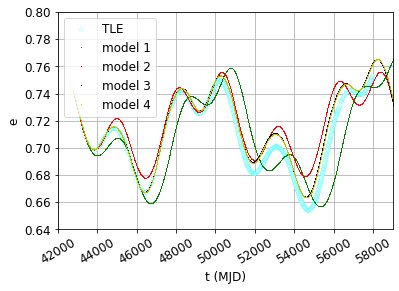}\\
\includegraphics[width=0.32\columnwidth]{te_tepropsempl_final_new_7.png}  \includegraphics[width=0.32\columnwidth]{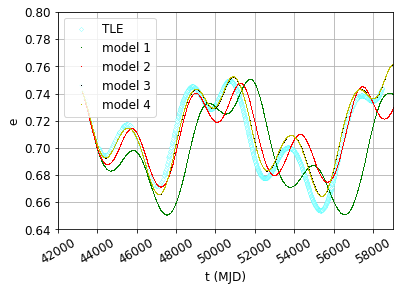} \includegraphics[width=0.32\columnwidth]{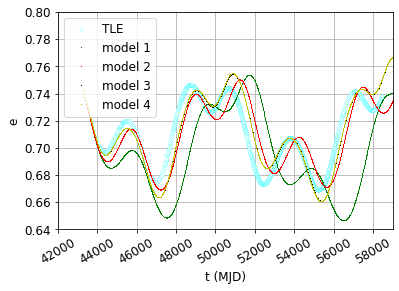}\\
\includegraphics[width=0.32\columnwidth]{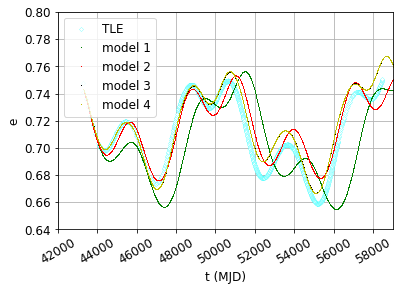}  \includegraphics[width=0.32\columnwidth]{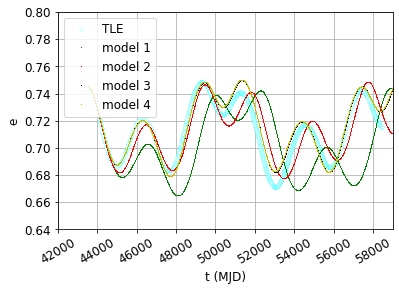} \includegraphics[width=0.32\columnwidth]{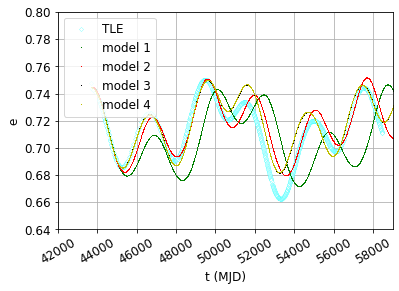}\\
\includegraphics[width=0.32\columnwidth]{te_tepropsempl_final_new_13.png}  \includegraphics[width=0.32\columnwidth]{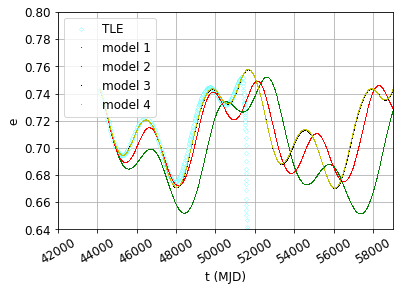} \includegraphics[width=0.32\columnwidth]{te_tepropsempl_final_new_15.png}
 \caption{}
    \label{fig:prop_semplici_1serie}
    \end{center}
\end{figure}

\noindent
In Fig.~\ref{fig:prop_semplici_1serie}, orbits \#1-15 are reported.

\begin{figure}[h!]
\begin{center}
\includegraphics[width=0.32\columnwidth]{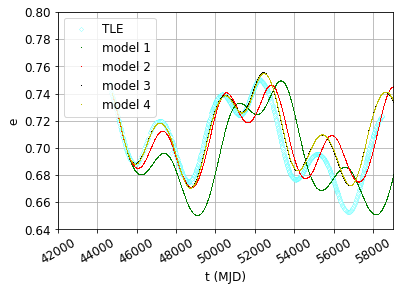}  \includegraphics[width=0.32\columnwidth]{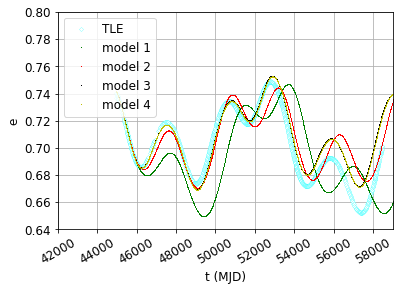} \includegraphics[width=0.32\columnwidth]{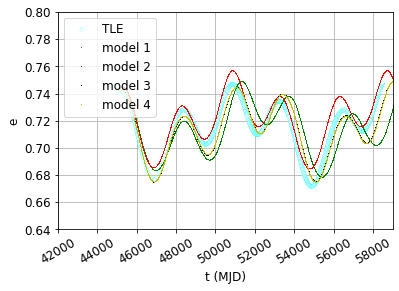}\\
\includegraphics[width=0.32\columnwidth]{te_tepropsempl_final_new_19.png}  \includegraphics[width=0.32\columnwidth]{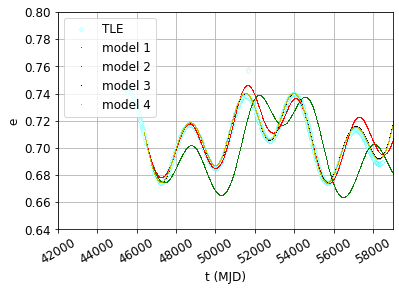} \includegraphics[width=0.32\columnwidth]{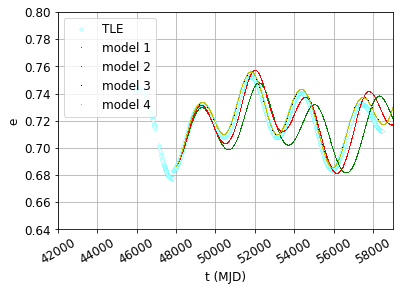}\\
\includegraphics[width=0.32\columnwidth]{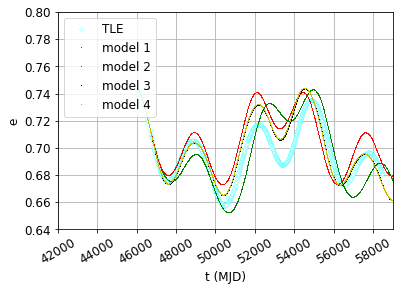}  \includegraphics[width=0.32\columnwidth]{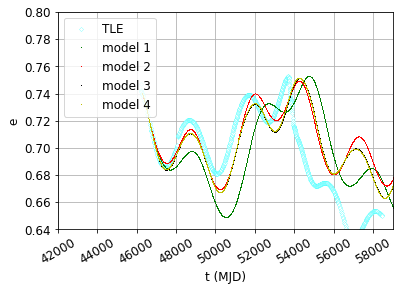} \includegraphics[width=0.32\columnwidth]{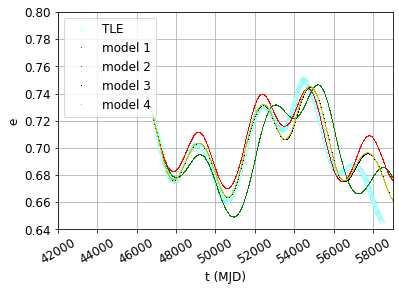}\\
\includegraphics[width=0.32\columnwidth]{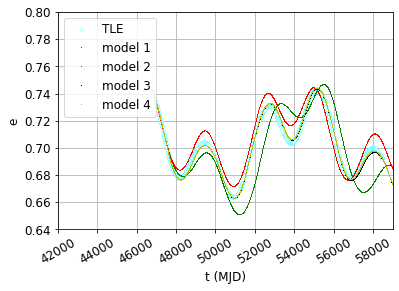}  \includegraphics[width=0.32\columnwidth]{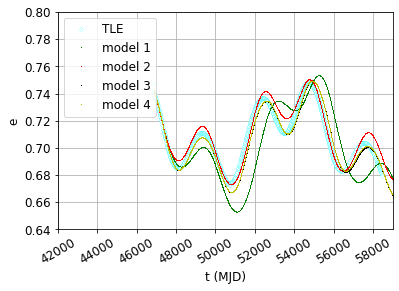} \includegraphics[width=0.32\columnwidth]{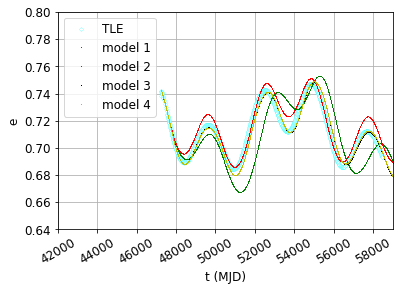}\\
\includegraphics[width=0.32\columnwidth]{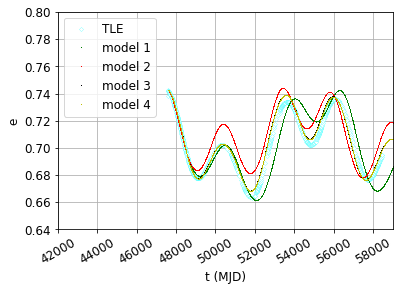}  \includegraphics[width=0.32\columnwidth]{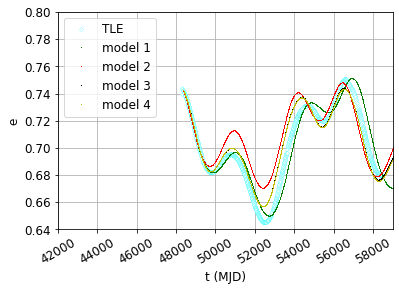} \includegraphics[width=0.32\columnwidth]{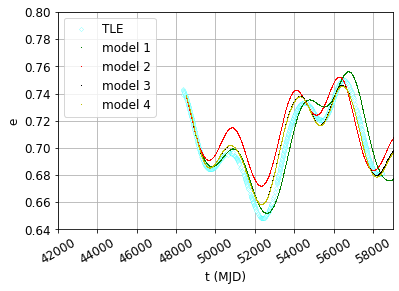}
 \caption{}
    \label{fig:prop_semplici_2serie}
    \end{center}
\end{figure}

\noindent
In Fig.~\ref{fig:prop_semplici_2serie}, orbits \#16-30 are reported.

\begin{figure}[h!]
\begin{center}
\includegraphics[width=0.32\columnwidth]{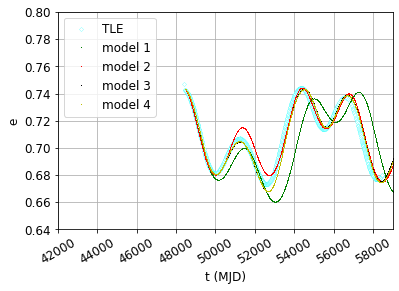}  \includegraphics[width=0.32\columnwidth]{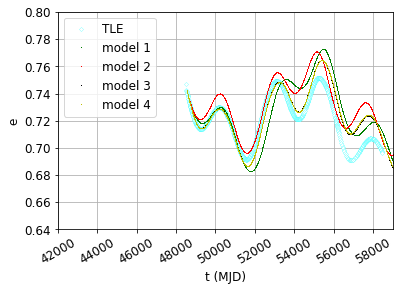} \includegraphics[width=0.32\columnwidth]{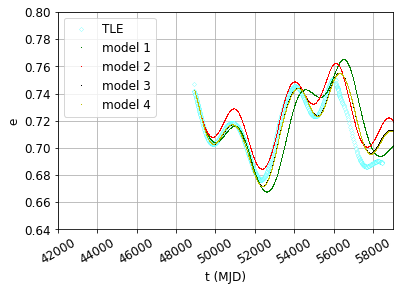}\\
\includegraphics[width=0.32\columnwidth]{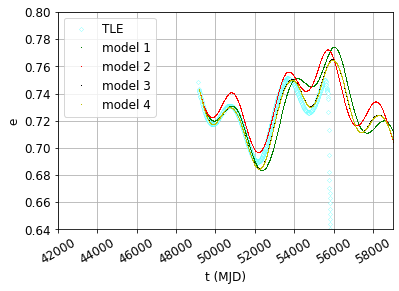}  \includegraphics[width=0.32\columnwidth]{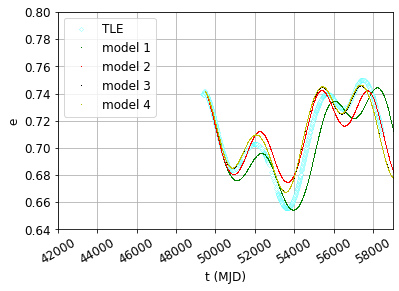} \includegraphics[width=0.32\columnwidth]{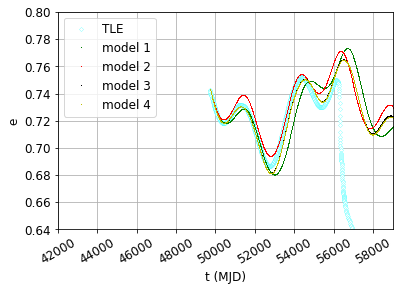}\\
\includegraphics[width=0.32\columnwidth]{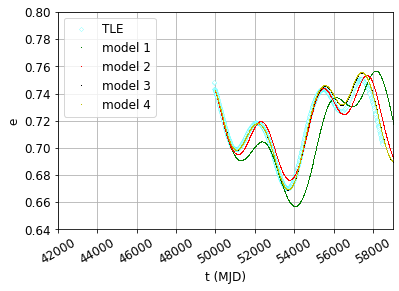}  \includegraphics[width=0.32\columnwidth]{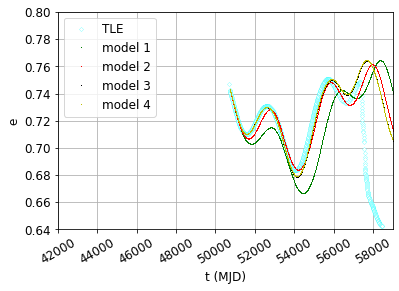} \includegraphics[width=0.32\columnwidth]{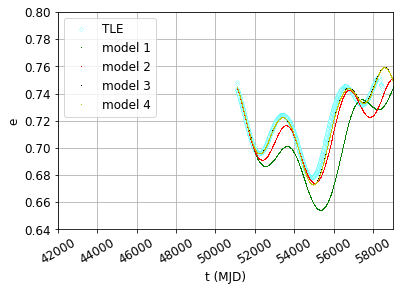}\\
\includegraphics[width=0.32\columnwidth]{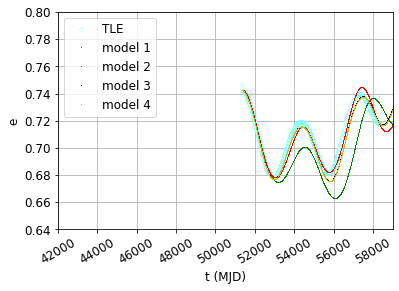}  \includegraphics[width=0.32\columnwidth]{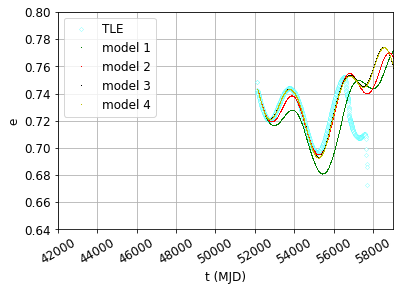} \includegraphics[width=0.32\columnwidth]{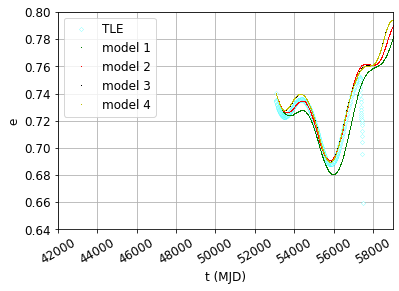}
 \caption{}
    \label{fig:prop_semplici_3serie}
    \end{center}
\end{figure}

\noindent
In Fig.~\ref{fig:prop_semplici_3serie}, orbits \#31-42 are reported.

\end{document}